\documentclass[lettersize,journal]{IEEEtran}
\usepackage{amsmath,amsfonts}
\usepackage{algorithm}
\usepackage{array}
\usepackage[caption=false,font=normalsize,labelfont=sf,textfont=sf]{subfig}
\usepackage{textcomp}
\usepackage{stfloats}
\usepackage{url}
\usepackage{verbatim}
\usepackage{graphicx}
\usepackage{cite}
\usepackage{tabularx,makecell}
\usepackage{xcolor}
\usepackage{multirow}
\usepackage{algpseudocode}

\newcommand\GOLDENFIRINGRECORD{\mathit{golden\_record}}
\newcommand\FIRINGRECORD{\mathit{record}}
\newcommand\NORMALSEQ{\mathit{normal\_seq}}
\newcommand\ENDTRANS{\mathit{end\_trans}}
\newcommand\PREVTRANS{\mathit{prev\_state}}
\newcommand\NEWTRANS{\mathit{state}}
\newcommand\LASTTRANS{\mathit{last\_trans}}

\newcommand\ELEMENT{\mathit{element}}
\newcommand\SEQ{\mathit{seq}}

\newcommand\NTP{\mathit{N_{TP}}}
\newcommand\DR{\mathit{DR}}
\newcommand\NOE{\mathit{N_{OE}}}

\newcommand\NINJ{\mathit{N_{inj.}}}

\newcommand\NREGS{\mathit{N_{regs}}}
\newcommand\FAULTDETECTED{\mathit{fault\_detected}}

\newcommand\NINPUT{\mathit{N_{input}}}
\newcommand\LATENCY{\mathit{Lat}}
\newcommand\DRTO{\mathit{DR\_TO}}
\newcommand\AREA{\mathit{Area}}
\newcommand\SEQUENCES{\mathit{Seqs}}

\begin{document}

\title{In-Situ Hardware Error Detection Using Specification-Derived Petri Net Models and Behavior-Derived State Sequences}


\author{Tomonari Tanaka, Takumi Uezono, Kohei Suenaga, Masanori Hashimoto

\thanks{This study was partially supported by JSPS KAKENHI Grant Number 24H00073, and JST SPRING, Grant Number JPMJSP2110.}

\thanks{T. Tanaka, K. Suenaga, and M. Hashimoto are with the Department of Communications and Computer Engineering, Kyoto University, Kyoto, 606-8501, Japan.}
\thanks{T. Uezono is with Production Engineering and MONOZUKURI Innovation Center, Center for Sustainability, Research and Development Group, Hitachi, Ltd., Yokohama 244-0817, Japan}

}

\markboth{}
{Shell \MakeLowercase{\textit{et al.}}: A Sample Article Using IEEEtran.cls for IEEE Journals}


\maketitle

\begin{abstract}
In hardware accelerators used in data centers and safety-critical applications, soft errors and resultant silent data corruption significantly compromise reliability, particularly when upsets occur in control-flow operations, leading to severe failures. To address this, we introduce two methods for monitoring control flows: using specification-derived Petri nets and using behavior-derived state transitions.
We validated our method across four designs: convolutional layer operation, Gaussian blur, AES encryption, and a router in Network-on-Chip. Our fault injection campaign targeting the control registers and primary control inputs demonstrated high error detection rates in both datapath and control logic.
Synthesis results show that a maximum detection rate is achieved with a few to around 10\% area overhead in most cases.
The proposed detectors quickly detect 48\% to 100\% of failures resulting from upsets in internal control registers and perturbations in primary control inputs.
The two proposed methods were compared in terms of area overhead and error detection rate. By selectively applying these two methods, a wide range of area constraints can be accommodated, enabling practical implementation and effectively enhancing error detection capabilities.
\end{abstract}


\section{Introduction}
\IEEEPARstart{H}{ardware} accelerators that process tasks like image processing and AI inference are increasingly used in various domains, with heightened demand for reliability in safety-critical applications such as autonomous driving and medical devices~\cite{Rech2024, Jha2019, Fausti2019, Tanaka2021}. 
Silent data corruption in data centers accommodating hardware accelerators draws significant attention~\cite{Bittel2024,Keller2021,Konno2024}. 
The causes of silent data corruption, such as bit flips in memory, are due to several factors, including cosmic rays, temperature variations, voltage fluctuations, and aging effects~\cite{Hashimoto2020,Muslim2015,Enkele2024}.
Among these factors, soft errors due to cosmic rays in terrestrial and space environments are the primary causes of silent data corruptions, and pose significant reliability concerns for these accelerators across their lifetime~\cite{Lopes2018, Du2019}.

Available methods for evaluating hardware accelerator reliability against soft errors include irradiation experiments and fault injection. The former uses actual radiation to deliver accurate assessments but is limited by time and facility availability. Conversely, fault injection experiments are more flexible, allowing for the controlled injection of bit upsets over time and space, with an option to repeat evaluations as needed.
Fault injection targeting hardware accelerators has demonstrated that control registers related to control flow are especially vulnerable to bit upsets~\cite{Hoefer2023,Sabogal2021}.

Fault-tolerant methods like instruction redundancy in software and hardware lockstep are proposed to detect soft errors impacting control flow~\cite{Bohman2019, Iturbe2016, de2018}. For hardware, Dual Modular Redundancy (DMR) and Triple Modular Redundancy (TMR) are often used, supplemented by application-specific strategies~\cite{Libano2019, Bertoa2023}. However, even TMR has been reported as ineffective against single points of failure, such as shared I/O~\cite{Cannon2020}, indicating they cannot cope with input failures.
Additionally, existing error detection techniques often target the data and control flows of specific applications~\cite{Li2020, Hari2022, Younis2020}.
As a more generalized error detection method, P.~Taatizadeh and N.~Nicolici proposed an assertion-based bit-flip detection technique and evaluated their method using three benchmark circuits~\cite{Pouya2016}. However, a comprehensive evaluation for practical applications has not been conducted. 

Implementing error detectors in hardware is an effective strategy for diagnosing hardware throughout its lifetime, which enables quick error detection. M.~Boule et al. 
have proposed a method to integrate a dedicated programmable region for error detectors within Application-Specific Integrated Circuits (ASICs)~\cite{Boule2007}, allowing flexible adaptation to specific purposes. In such a configuration, error detectors can be customized according to the objectives of error detection.
Nevertheless, devising new error detection methods that achieve high fault detection rates with minimal area overhead remains a critical challenge in such approaches.



The primary goal of this work is to expand the options for detecting control flow errors, enabling designers to accommodate various design constraints while enhancing error detection capabilities for both internal and input failures. To this end, we propose two in-situ error detection methods.
The first method is based on Petri nets, which are constructed to represent the control-flow specifications. Our preliminary work on this Petri-net-based error detection was reported with fewer design examples in \cite{Tanaka2025}. This approach employs multiple compact Petri nets to detect most control-flow perturbations caused by both bit flips and input failures, as well as any resulting incorrect datapath outputs. These Petri nets can be integrated into hardware for error detection with minimal overhead, ensuring no false error detection in error-free operations.
In this work, we also introduce a second approach to improve applicability to circuits with diverse characteristics and ensure adaptability to a wide range of area constraints. This second approach involves defining and constructing state sequences to diagnose the state of target hardware accelerators. By carefully selecting monitored signals at different hierarchical levels, designers can balance the trade-off between error detection rate and area overhead. Thus, these approaches achieve a high error detection rate across a broad spectrum of area constraints.
Key contributions of this work are summarized as follows:
\begin{itemize}
    \item Establishing generalized error detection methods targeting both register bit-flips and input perturbations.
    \item Applying the proposed methods to practical designs that can be used in real-world environments, achieving a high error detection rate.
    \item Implementing detectors under various area constraints to explore the trade-off between error detection rate and area overhead.
\end{itemize}

The remainder of this paper is organized as follows. Section~\ref{sec_related_works} surveys related work, and Section~\ref{sec_propesed_method} presents the proposed methods. Section~\ref{sec_design_example} describes the application of these methods to design examples, while Section~\ref{sec_experimental_results} details the experimental results. Finally, Section~\ref{sec_conclusion} concludes the paper.

\section{Related work 
}
\label{sec_related_works}
We aim to develop a method for detecting hardware failures efficiently due to soft errors by monitoring hardware control flow. In reviewing relevant research, several strategies emerge:

\textit{Hardware-based Error Detection in Datapaths:} Z.~Zhu and B.~C.~Schafer proposed periodic monitoring using pre-acquired golden values for implementations based on high-level synthesis (HLS)\cite{Zhu2020}. W.~Li et al. also proposed a periodic error detection method that employs golden data during idle times in Convolutional Neural Network (CNN) accelerators, allowing error detection without affecting the processing cycle counts\cite{Li2020}. However, applying this method to a wide range of applications still poses a challenge. Since hardware-based error detection methods are generally less flexible than software-based ones, evaluation across multiple applications is valuable.

\textit{Software Techniques for Transient Error Mitigation:} S.~K.~S.~Hari et al. proposed low-cost error detectors based on vulnerability identification and redundant programming~\cite{Hari2012}. M.~Bohman et al. proposed instruction duplication techniques using a customized compiler to enhance resilience against soft errors~\cite{Bohman2019}. M.~Didehban et al. improved the insertion mechanism of redundant instructions to enhance error detection performance against transient faults~\cite{Didehban2024}.  Although instruction duplication mitigates silent data corruptions from datapath faults, control flow-related faults like hang-ups remain challenging, with TMR showing no improvement.

\textit{Fault Injection and Reliability Evaluation:} M.~H.~Ahmadilivani et al. explored the vulnerability of CNN accelerators by performing fault injections on neuron outputs, where bit-flips significantly impact the results~\cite{Ahmadilivani2023}. Z.~Chen et al. identified sensitive bits in binary data within machine learning applications to improve the efficiency of fault injection campaigns~\cite{Chen2019}. While these methods effectively identify vulnerable bits, they still face challenges in achieving efficient hardware failure detection.

\textit{Soft Error Impact Mitigation:} L.~Chen and M.~Tahoori proposed an approach that selectively protects crucial registers in control and data flows via HLS to reduce error probability, though it prioritizes error probability reduction over detection~\cite{Liang2014}. S.~T.~Fleming and D.~B.~Thomas proposed a tool that extracts and protects control flow from C-language descriptions for HLS, mainly detecting errors affecting execution times rather than computational accuracy~\cite{Fleming2016}.
M.~J.~Cannon et al. evaluated the performance of TMR techniques through irradiation experiments~\cite{Cannon2020}. In normal TMR configurations, a shared primary input is a single point of failure, implying that TMR cannot mitigate errors when inputs are affected by soft errors, thus compromising the benefits of redundancy.

\textit{Machine Learning for Error Detection:} N.~ Nosrati et al. proposed a machine learning-based method to monitor crucial signals related to control flow in microprocessors~\cite{Nosrati2022}. Despite its general applicability, this method risks misidentifying fault-free operations as faulty, which has been confirmed as a significant problem in ~\cite{Tanaka2025}.

\textit{Assertion-based Error Detection:} P.~Taatizadeh and N.~Nicolici proposed a method that utilizes assertions automatically generated from Hardware Description Language (HDL) and simulation traces~\cite{Pouya2016}. Subsequently, they incorporated a SAT solver to derive more accurate invariants, such as assertions~\cite{Pouya2017}. However, these methods have been evaluated using benchmarks rather than practical applications, and the long computational time has also been reported. As demonstrated in Section~\ref{sec_comparison_with_related_work}, the adaptation to practical circuits remains a challenge.

Thus, expanding error detection methods for control flow errors while improving design efficiency and reducing overhead remains a significant challenge, despite strong demand.

\section{Proposed method}\label{sec_propesed_method}
We aim to establish error detection methods that address not only single-event upsets within a target module but also errors propagated from upstream circuits, which simple register duplication cannot detect. Table~\ref{tab_overview_of_methods} overviews our two proposed methods. The Petri-net-based approach directly monitors selected design specifications, while the state-sequence-based approach detects abnormal transitions using state sequences obtained from application execution. Both methods can be implemented as dedicated hardware on programmable logic, making them adaptable to diverse requirements.

Our primary goal is to offer additional options rather than merely surpassing existing error detection techniques. These methods can also be selectively employed or combined with others, depending on design needs. Section~\ref{sec_experimental_results} presents a quantitative comparison of detection performance and area overhead through fault-injection experiments.

\begin{table*}
\centering
\small
  \caption{Overview and qualitative comparison of the proposed methods.}
\label{tab_overview_of_methods}
  \begin{tabularx}{\linewidth}{llXl}
    \hline
    Approach & Source & Description & Implementation\\
    \hline
    Petri-net-based &
    Specification& 
    Multiple small Petri nets are generated to monitor control flow-related specifications, ensuring no false positives in error detection during normal operations. For efficient implementation, a subset of Petri nets are selected, considering trade-off between area-overhead and error detection rate.&
    \makecell[l]{Hardware\\(Dedicated programmable logic)}
    \\
    \hline
    State-sequence-based &
    Behavior& 
    Normal state sequences are obtained from error-free RTL simulations to detect abnormal state sequences.
    To ensure optimal implementation and prevent the explosion of state sequences, normal state sequences are derived from the behavior resulting from normal input patterns.
    &
    \makecell[l]{Hardware\\(Dedicated programmable logic)}
    \\
    \hline
  \end{tabularx}
\end{table*}

\subsection{Petri-net-based error detection}
\begin{figure}[!tb]
  \centering
  \includegraphics[width=\columnwidth]{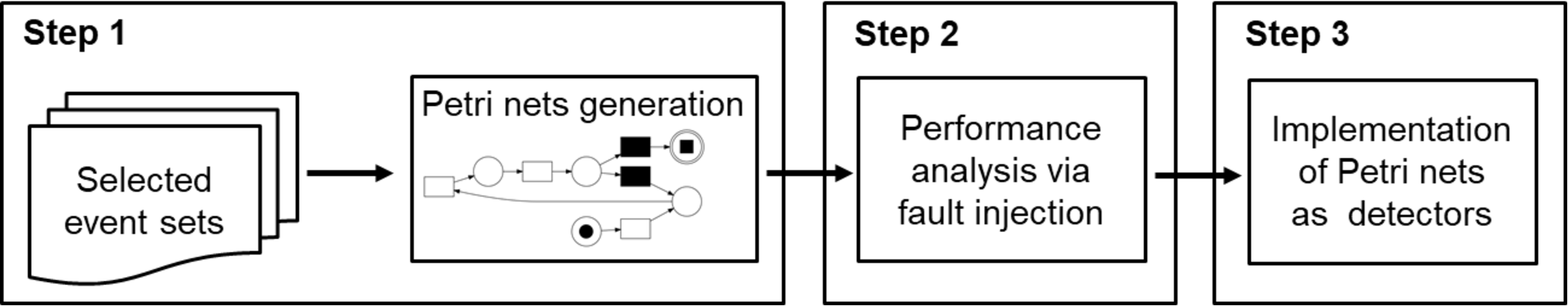}
  \vspace*{-0mm}
  \caption{Proposed method from generating Petri nets to implementing those as detectors. 
  }
  \label{fig_method}
\end{figure}

Fig.~\ref{fig_method} shows the proposed Petri-net-based error detection consisting of three steps: (1) generating Petri nets from specifications, (2) evaluating their fault detection performance, and (3) selecting Petri nets based on area and fault detection performance and implementing the selected ones as detectors. 
This Petri-net-based method assumes the existence of a specification document that fully describes signal changes within the hardware, which is typical in industrial designs, especially reliability-critical hardware.

\subsubsection{{Petri nets}}
Before explaining each step in detail, we give the definition of Petri nets used in this work.
  A Petri net is a mathematical model used to describe the behavior of a discrete event system.
  Structurally, a Petri net $S$ is a directed bipartite graph whose vertices are divided into two sets---\emph{places} $P$ and \emph{transitions} $T$---connected by directed edges $E$.
  Each place $p \in P$ keeps a non-negative number of \emph{tokens}.
  A state of a Petri net is represented by a function $M : P \rightarrow \mathbb{N}$, where $M(p)$ denotes the number of tokens in place $p$.
  We call such a function $f$ a \emph{marking} of the Petri net.
  The initial marking of a Petri net is denoted by $M_0$.
  A transition $t \in T$ is said to be \emph{enabled} if every place $p$ such that $(p,t) \in E$ (i.e., $p$ is an input place of $t$) contains at least one token.
  A Petri net changes its state by \emph{firing} one of the enabled transitions; once an enabled transition $t$ is fired, it consumes one token from each input place $p_{I}$ (i.e., $(p_I, t) \in E$) and produces one token to each output place $p_O$ (i.e., $(t, p_O) \in E$).
  In this way, a Petri net models a sequence of discrete events as a sequence of transition firing.

\begin{figure}
  \centering
  \includegraphics[width=.9\linewidth]{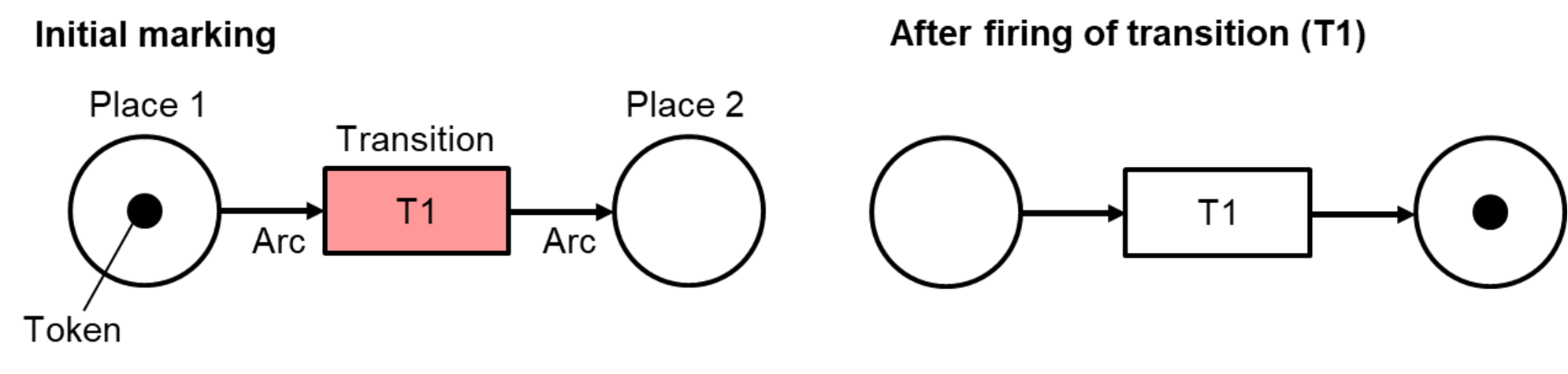}
  \vspace*{-0mm}
  \caption{Simple Petri net.} 
  \label{fig_petri_net}
\end{figure}

  Fig.~\ref{fig_petri_net} shows a simple Petri net with two places and one transition.
  In the initial marking, Place 1 contains one token, while Place 2 is empty.
  The transition $\mathrm{T1}$, whose input is Place 1 and output is Place 2, is enabled since Place 1 contains a token.
  After $\mathrm{T1}$ fires, the token in Place 1 is consumed, and a token is produced in Place 2.  

Petri nets have been widely used in studies related to hardware security~\cite{Guechi2023,Patzina2010,Song2021,Wang2007}. However, there has been relatively little research on the efficient implementation of Petri nets as runtime checkers for soft errors.
The following sections explain each step, from the construction of Petri nets to their implementation.

\subsubsection{Step 1: Extract event sets from specification and generate their corresponding Petri nets}
In the specification document, we categorized specific signal changes into four types, interpreting them as individual events.
Types 1 and 2 involve changes in the value of a specific signal; Type 1 includes any change, while Type 2 focuses on changes to specific values. 
Types 3 and 4 relate to the $i$-th change of the signal value, where $i$ is a predetermined value. Type 3 captures any $i$-th changes, whereas Type 4 is restricted to $i$-th changes of specific values.
For example, considering a status signal indicating two states (\textit{S1}, \textit{S2}), changing the signal value represents a state transition. A simple state transition is Type~1, while a transition to a specific state (e.g., \textit{S2}) is Type~2. The i-th state transition is Type~3, and the i-th transition to a specific state (e.g., $i$-th \textit{S1}) is Type~4.
Considering the implementation, the allocation type of an event is relevant to the hardware resources required for its observation. For instance, observing a Type 4 event requires more hardware resources than observing a Type 1 event to count the number of transitions. We finally consider the balance between hardware resources and error detection performance, which will appear in Section~\ref{sec_peformance_pn}.

Given our focus on monitoring the control flow of hardware accelerators, we extract an event set that meets a specific condition: \emph{There must be at least two target events, and their occurrence order must remain consistent across multiple executions.}
For instance, this condition is met if a monitored specification includes events A, B, and C; and if
these events consistently occur in the order of A, B, and then C during correct executions.
In this paper, we manually identify event sets. Meanwhile, large language models (LLMs) or automated assertion techniques, e.g., \cite{Zhang2017, Germiniani2022, Fang2024}, may help.

Next, we generate Petri nets corresponding to individual event sets. Each event is assigned to a transition, and a Petri net is constructed to represent the sequence of these event occurrences. To enhance monitoring capabilities, multiple event sets and their Petri nets are generated. 
We use Petri nets to handle complex control flows in anticipation of future demands. However, within the scope of this paper, alternatives such as automata may also be applied.

\subsubsection{Step 2: Evaluate error detection performance}
\paragraph{Simulating Petri-net-based error detection}

Error detection with Petri nets is achieved by monitoring the sequence of transition firings.
If an abnormal firing sequence (including the firing of an incorrect final transition) that is not defined by the Petri net is detected, it is considered that an error has been detected.
The focus here is on monitoring control flow; thus, fault injections simulate bit flips in registers responsible for control flow, as well as in primary control inputs.
Detection of output errors is indicated by abnormal transition firings, where the output error is a fault resulting in incorrect outcomes. The output errors are categorized either as incorrect computational results, namely silent data corruption defined by deviations from correct values, or as abnormal terminations of processing, such as timeouts or premature termination.
These faults can affect both the datapath and control logic.
When a Petri net detects an error upon the occurrence of an output error, it is considered true error detection.

\paragraph{Metrics of error detection performance}\label{sec_petri_net_metrics_error_detection}

We evaluate the error detection performance of Petri nets using two metrics: error detection rate and error detection latency. 
When an error is detected in the Petri net, it is expected to indicate that the hardware produces incorrect outputs. Therefore, if the Petri net detects an error and the hardware outputs a fault, it is counted as a true positive error detection~($\NTP$). Then, the error detection rate~($\DR$) is defined as $\DR=\NTP / \NOE$, where $\NOE$ is the number of output error occurrences.

Additionally, we evaluate the latency of error detection using Petri nets. Latency ($\LATENCY$) is defined as the average number of clock cycles between the injection of a fault and the Petri net detecting the error. In scenarios where the process fails to complete and results in a timeout, the Petri net may only detect the incorrect final transition. Given that the practical timeout duration is not fixed, the latency of error detection becomes ambiguous. Consequently, the proportion of such error detections is calculated as $\DRTO$, which is a subset of the $\DR$. A smaller $\LATENCY$ and a lower $\DRTO$ are indicative of better error detection capabilities. 


\subsubsection{Step 3: Select and implement Petri nets as error detectors}
\begin{figure}[!tb]
  \centering
  \includegraphics[width=.6\linewidth]{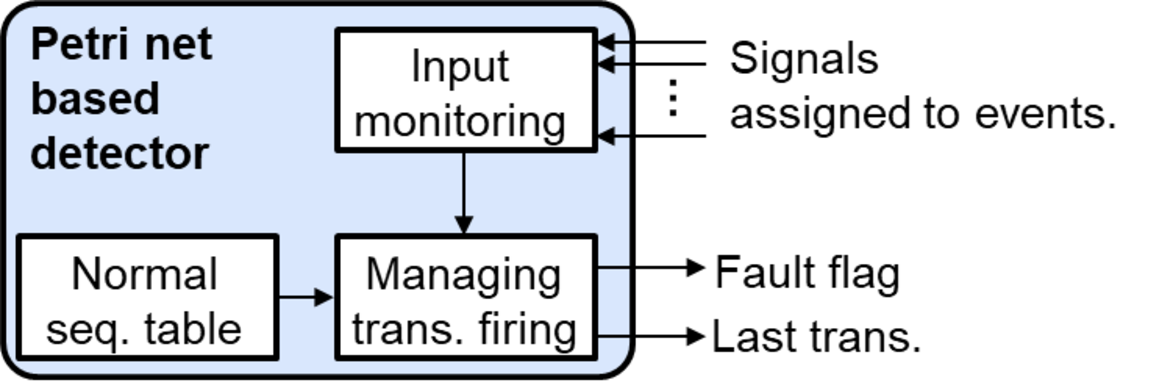}
  \vspace*{-0mm}
  \caption{Architecture of Petri-net-based error detector.
  }
  \label{fig_pn_detector}
  \vspace*{-0mm}
\end{figure}

This step involves implementing Petri nets as error detectors to monitor hardware failures in real time. Fig.~\ref{fig_pn_detector} illustrates the architecture of the Petri-net-based error detector, consisting primarily of three parts: input monitoring, managing transition firing, and the normal sequence table. The input to the error detector consists of the signal lines assigned to events. These input signals are monitored by the input monitoring module to detect changes in signals and the associated events. Event occurrences are defined as the transition firing in the Petri nets. The transition-firing management module monitors the firing sequence of transitions based on the normal sequence table. When the module detects transitions firing in abnormal sequences, it asserts the fault flag (Fault flag) to indicate error detection. Additionally, the transition-firing management module constantly outputs the last-fired transition (Last trans.), enabling the detection of abnormal process terminations.

When maximizing detection rate (DR) with an area overhead constraint, we find the combination of detectors that achieves the highest DR while satisfying the area constraint. When minimizing area with a DR constraint, we identify the detector combination with the minimum area overhead while meeting the DR constraint.

\subsection{State-sequence-based error detection}
\subsubsection{Step 1: Acquiring normal state sequences}\label{sec_acq_normal_state_seq}

\begin{figure}
  \centering
  \includegraphics[width=\linewidth]{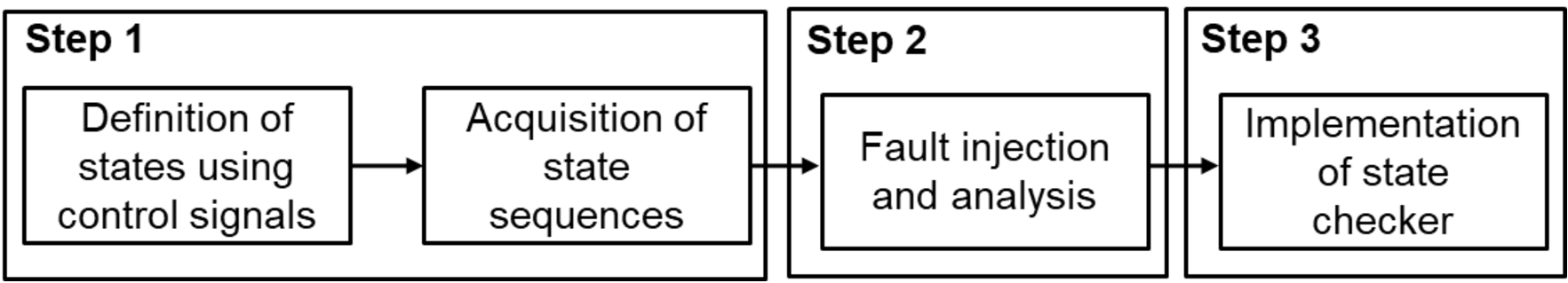}
  \caption{Proposed error detection method using state sequences.}
  \label{fig_proposed_method_seq}
\end{figure}

\begin{figure}
  \centering
  \includegraphics[width=0.65\linewidth]{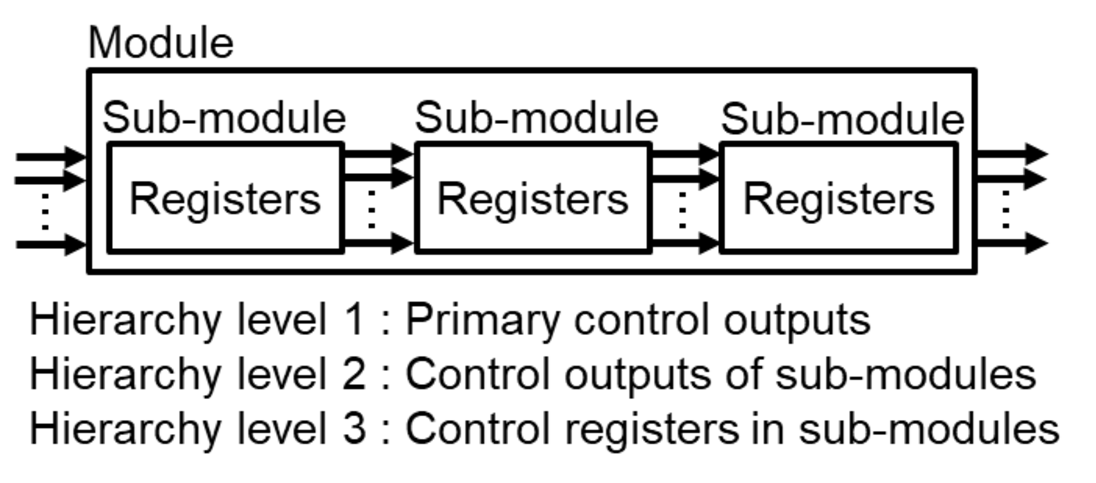}
  \caption{Monitored signals at different hierarchical levels. 
  }
  \label{fig_hierarchy}
\end{figure}

Fig.~\ref{fig_proposed_method_seq} illustrates the proposed error detection method based on state sequences. First, the state is defined using control signals at various hierarchical levels of the target circuits, as shown in Fig.~\ref{fig_hierarchy}. By combining values from multiple signals, this new state representation accommodates diverse circuit behaviors. Monitored signals span three hierarchical levels: (1) the primary outputs of the target module, (2) the primary outputs of sub-modules, and (3) control registers within the target module. Since the goal is to monitor control flow, only control-related signals are selected. Generally, higher levels feature fewer monitored signals, reducing the hardware area of the error detector. Results for a specific example appear in Section~\ref{sec_design_example}.
Table~\ref{tab_monitored_bit_types} shows the bit-selection types for monitoring. While monitoring all bits can enhance error detection, it demands more hardware. To address various area constraints, we also consider the most significant bits (MSBs), MSBs in the utilized bit range, and least significant bits (LSBs).

\begin{table}[t]
\centering
\small
  \caption{Monitored bit selection types }
 \label{tab_monitored_bit_types}
  \begin{tabular}{cl}
    \hline
    Type & Monitored bits \\
    \hline
     1 & All bits \\
    \hline
     2 & Most significant bits (MSBs) \\
    \hline
     3 & MSBs within the
used bit range \\
    \hline
     4 & Least significant bits (LSBs)\\
    \hline
  \end{tabular}
\end{table}

Second, the state sequence of transitions is captured via RTL simulation. Algorithm~\ref{alg_normal_seq} explains how to obtain normal state sequences from $\GOLDENFIRINGRECORD$, which stores chronological state transitions generated by the golden simulation. In the \textit{for} loop (starting at line 4), 
$\GOLDENFIRINGRECORD$ is scanned from the beginning, logging the state of each clock cycle in 
$\ELEMENT$. Hence, the loop captures a continuous record of circuit behavior, from which 
$\ELEMENT$ is assigned to 
$\NEWTRANS$ as the last-fired state while 
$\PREVTRANS$ holds the previous state. Together, 
$\PREVTRANS$ and 
$\NEWTRANS$ form the two-length sequence 
$\SEQ$. Ultimately, Algorithm~\ref{alg_normal_seq} returns 
$\NORMALSEQ$, the normal state sequences, and 
$\ENDTRANS$, the last-fired transition.

Collecting all possible state sequences can be expensive, as circuits often handle numerous input patterns. However, since we assume a specific application, the number of input patterns producing different state sequences can be restricted.

\begin{figure}[t]
\begin{algorithm}[H]
    \caption{Obtaining normal state sequences from a golden simulation.}
    \label{alg_normal_seq}
    \begin{algorithmic}[1]
    \Function {acq\_normal\_seq}{$\GOLDENFIRINGRECORD$}
        \State $\mathit{prev\_state} \gets ""$
        \State $\mathit{normal\_seq} \gets []$
        \ForAll{$\mathit{element} \gets \mathit{golden\_firing\_record}$}
            \State $\mathit{state} \gets \mathit{element}$
            \State $\mathit{seq} \gets (\mathit{prev\_state}, \mathit{state})$
            \If {$\mathit{seq} \not \in \mathit{normal\_seq}$}
                \State $\mathbf{add}(\mathit{normal\_seq}, \mathit{seq})$
            \EndIf
            \State $\mathit{prev\_state} \gets \mathit{state}$
        \EndFor
        \State $\mathit{end\_trans} \gets \mathit{new\_trans}$
        \State \Return $\mathit{normal\_seq}, \mathit{end\_trans}$
    \EndFunction
    \end{algorithmic}
\end{algorithm}
\end{figure}

\begin{figure}[t]
\begin{algorithm}[H]
    \caption{Error detection using normal state sequences.}
    \label{alg_detect_fault}
    \begin{algorithmic}[1]
    \Function {detection}{$\FIRINGRECORD, \NORMALSEQ, \ENDTRANS$}
        \State $\PREVTRANS \gets ""$
        \State $\FAULTDETECTED \gets \mathit{False}$
        \ForAll {$\ELEMENT \gets \FIRINGRECORD$}
            \State $\NEWTRANS \gets \ELEMENT$
            \State $\SEQ \gets (\PREVTRANS, new\_trans)$
            \If {$\SEQ \not \in \NORMALSEQ$}
                \State $\FAULTDETECTED \gets \mathit{True}$
            \EndIf
            \State $\PREVTRANS \gets \NEWTRANS$
        \EndFor
        \State $\LASTTRANS \gets \NEWTRANS$
        \If {$\LASTTRANS \neq \ENDTRANS$}
            \State $\FAULTDETECTED \gets \mathit{True}$
        \EndIf
        \State \Return $\FAULTDETECTED$
    \EndFunction
    \end{algorithmic}
\end{algorithm}
\end{figure}

\subsubsection{Step 2: Error detection using normal state sequences}

Error detection performance is evaluated through fault injection simulations using the same metrics as the Petri-net-based method. Detection with normal state sequences is based on monitoring state transitions. If a transition not found in the normal sequences occurs, it is flagged as an error. Since the focus is on control flow, fault injections target bit flips in control-related registers and primary control inputs.

Algorithm~\ref{alg_detect_fault} details the error detection using normal state sequences. 
This process compares $\FIRINGRECORD$—a chronological record of state transition during a fault injection simulation—with $\NORMALSEQ$, obtained via Algorithm~\ref{alg_normal_seq}. 
The procedure is akin to Algorithm~\ref{alg_normal_seq}, but specifically, line~7 checks if $SEQ$ is absent from $\NORMALSEQ$. If absent, it is considered an abnormal state sequence, setting $\FAULTDETECTED$ to True. Additionally, if the last-fired transition, $\LASTTRANS$, does not coincide with the $\ENDTRANS$ from the golden simulation, $\FAULTDETECTED$ is set to $\mathit{True}$, confirming error detection.
Note that error detection using the $\LASTTRANS$ is optional, as the $\LASTTRANS$ cannot be specified when the circuit exhibits diverse behaviors in more general-purpose applications.

\subsubsection{Step 3: Implementation of state sequences checker}

\begin{figure}
  \centering
  \includegraphics[width=0.6\linewidth]{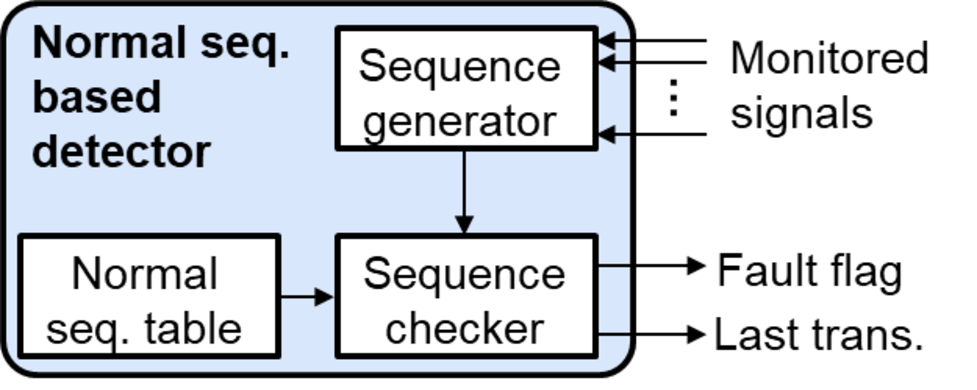}
  \caption{Normal-state-sequences-based error detector.}
  \label{fig_error_detector}
\end{figure}

This step implements error detectors using normal state sequences to monitor hardware failures in real time. Fig.~\ref{fig_error_detector} shows the detector architecture, consisting of three main parts: a sequence generator, a sequence checker, and a normal sequence table. The detector takes monitored signals as input.
The sequence generator creates a state by combining input signal values. The sequence checker monitors the state sequence against the normal sequence table. If it detects a sequence not in the table, it asserts a fault flag to signal an error. In the RTL description, all valid state sequences are explicitly defined; any sequence not listed is flagged as abnormal.
Additionally, a transition-firing management module outputs the last state continuously, enabling detection of abnormal terminations. The final detector is selected based on its error detection performance and area overhead.

\if
This step involves implementing error detectors using normal state sequences to monitor hardware failures in real time. Fig.~\ref{fig_error_detector} illustrates the architecture of the error detector, consisting primarily of three parts: a sequence generator, a sequence checker, and the normal sequence table. The input to the error detector are monitored signal.
The sequence generator constructs a state by combining the values of the input signals.
The sequence checker module monitors the state sequence based on the normal sequence table. When the module detects abnormal sequences, it asserts the fault flag (Fault flag) to indicate error detection.
In other words, in the actual RTL description, all normal state sequences are explicitly defined, and any state sequence that is not included among them is detected as an abnormal sequence.
Additionally, the transition-firing management module constantly outputs the last-state (Last state), enabling the detection of abnormal process terminations.
Finally, we choose the final detector based on its error detection performance and area overhead.
\fi

\section{Design example}
\label{sec_design_example}
We apply the proposed methods to four designs commonly found in application-specific accelerators, which we explain in the following sections.
\subsection{Convolutional layer computation}
Considering practical applications like autonomous driving, we first use a CNN accelerator~\cite{Huang2022} as an example. Since this accelerator accepts multiple sizes of input activation data and weight data, many operating patterns arise. In this work, we focus on one specific configuration to enhance error detection performance, assuming the programmable logic is reprogrammed for each. The target configuration is the first convolutional layer in a quantized LeNet-5 model, which employs INT8 precision and is trained on the MNIST dataset. The convolutional layer under test has $32\times32\times1$ input activation data and produces a $28\times28\times6$ output. Post-convolution, rectified linear unit (ReLU) activation functions are applied.

\begin{figure}
  \centering
  \includegraphics[width=.65\linewidth]{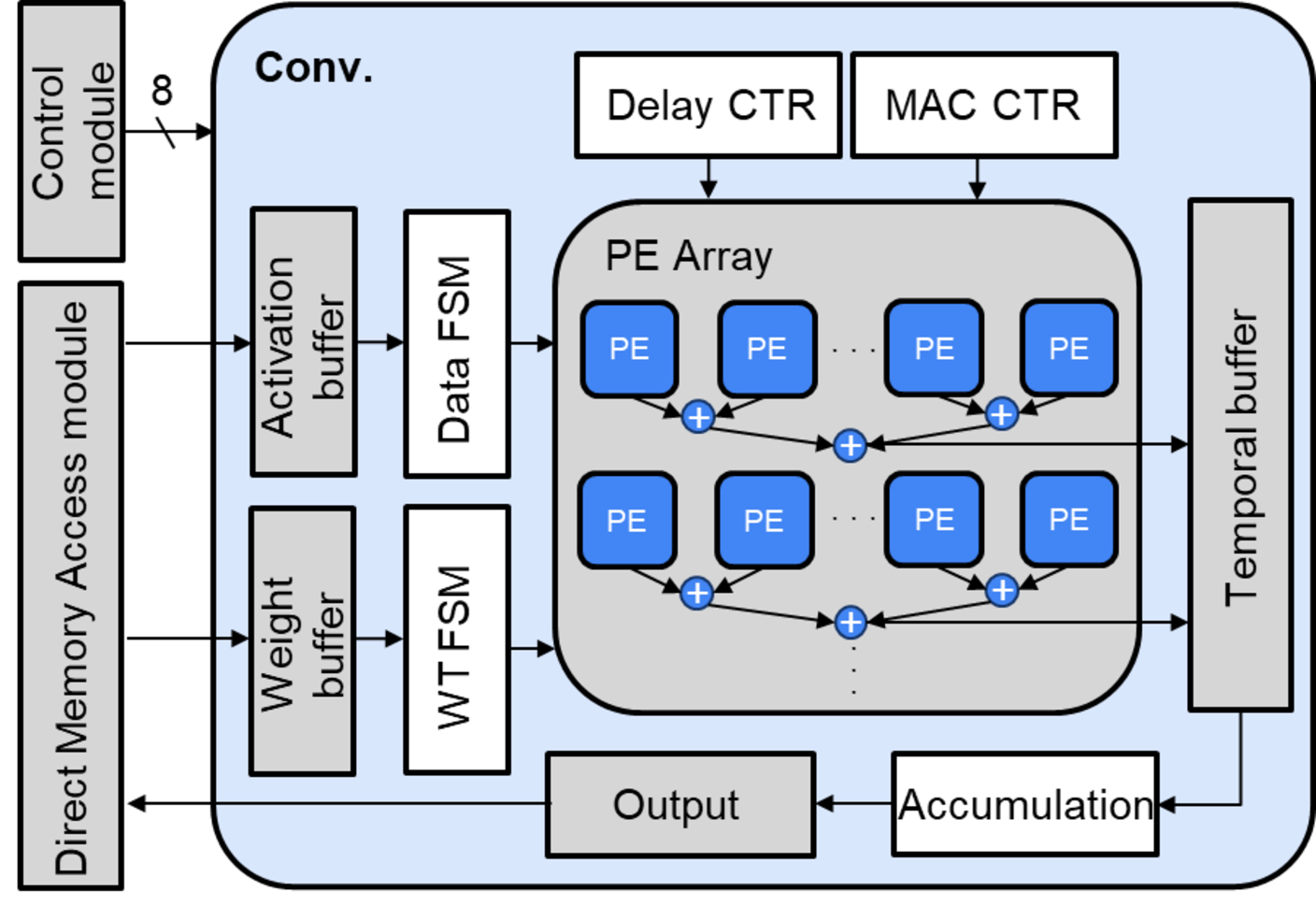}
  \caption{Architecture of CNN accelerator. White boxes are main modules.}
  \label{fig_cnn_acc}
\end{figure}

Fig.~\ref{fig_cnn_acc} illustrates the architecture of the CNN accelerator proposed in \cite{Huang2022}, which enables high data reuse and low latency performance. The convolution (Conv.) has primary control inputs from the control module. The input data, including weight data and activation data, is fed into the Weight buffer and Activation buffer, respectively. The WT FSM and Data FSM modules manage the input data from the buffers and control the main computation on the PE Array. The Delay CTR and MAC CTR also manage computation on the PE Array. The processed data in the PE Array is accumulated by the Accumulation module via the Temporal buffer and outputs the computation results to the Direct Memory Access module through the Output module.

\subsubsection{Petri-nets}
\begin{figure}
  \centering
  \includegraphics[width=.9\columnwidth]{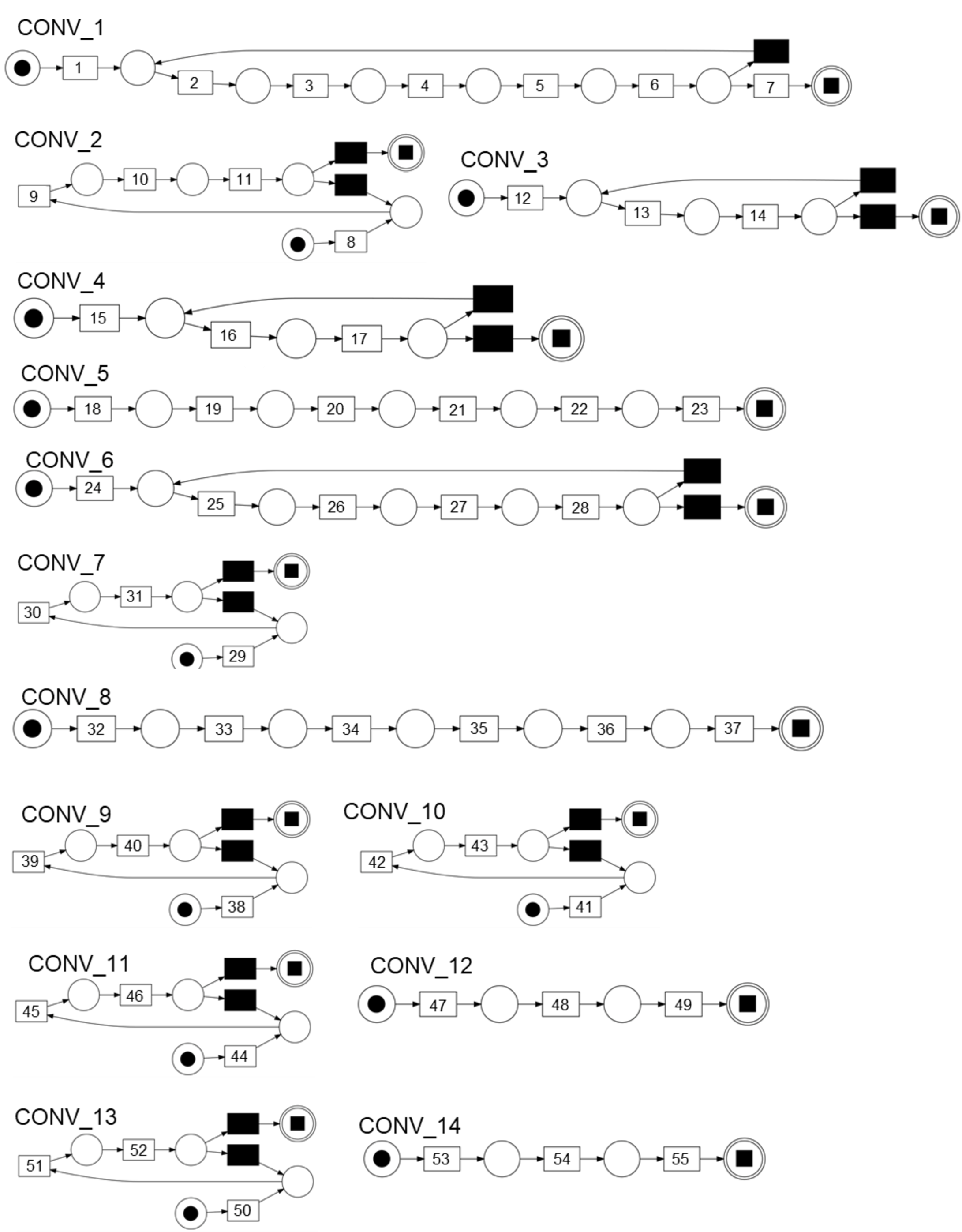}
  \caption{Petri nets for monitoring control-flow in Conv.}
  \label{fig_pn_lenet}
\end{figure}

To construct Petri nets that monitor the control flow, we first organize event sets from the specifications. Table~\ref{tab_construction_pn_cnn} details the event sets, their corresponding IDs for Petri nets, and the event assignment types used. Fig.~\ref{fig_pn_lenet} displays 14 generated Petri nets, each corresponding to an event set. Transition labels correspond to the event labels in Table~\ref{tab_construction_pn_cnn}. For example, the Petri net for CONV\_3 includes three defined events. This net features a path initiating with the first transition (12) and includes recurring transitions (13, 14). Black transitions indicate branches, with the bottom transition firing at the final loop, signifying the completion as the token moves to the rightmost place.

Logic synthesis for the CNN accelerator, along with 14 Petri nets, was conducted on the Kintex UltraScale FPGA with part xcku035-fbva900-1-i using the Vivado tool. The CNN accelerator used 20,936 LUTs and 17,739 FFs, and the 14 Petri nets used 2,050 LUTs and 1,045 FFs. These 14 Petri nets are just candidates, and a part of them will be selected as detectors. Additionally, we investigated the impact on the maximum operating frequency. While the maximum operating frequency in the original design was 119 MHz, it dropped to 118 MHz when the error detectors were added. The speed impact of Petri-net-based error detectors was negligible.

\begin{table*}
\small
  \caption{Monitored event sets, their corresponding IDs, and used event assignment types in Conv.}
  \label{tab_construction_pn_cnn}
\resizebox{\linewidth}{!}{
  \begin{tabularx}{\linewidth}{lXl}
    \hline
    ID & Event set with Event Label~(\#) & Type\\
    \hline
    CONV\_1 
    & Initiation of processing~(1), updating of horizontal counter for activation data~(2), retrieval of data cube~(3), completion of data cube retrieval~(4), writing to FIFO buffer~(5), completion of convolution calculation for the data cube~(6), conclusion of all computations~(7). 
    & 2\\
    \hline
    CONV\_2 & 
    Initiation of processing~(8), permission for data cube computation~(9), initiation of data cube computation~(10), updating of data cube~(11).
    & 2\\
    \hline
    CONV\_3 & 
    Initiation of processing~(12), retrieval of activation data from a specific position~(13), updating of vertical counter for activation data~(14).
     & 2,3\\
    \hline
    CONV\_4 & 
    Initiation of processing~(15), permission for weight data retrieval~(16), retrieval of weight data~(17).
     & 2,3\\
    \hline
    CONV\_5 & 
    Setting of input channel number~(18), setting of output channel number~(19), initiation of processing~(20), state change for processing~(21), computation of a specific data cube~(22),
    state change for completion~(23).
     & 2,3\\
    \hline
    CONV\_6 & 
    Initiation of processing~(24), retrieval of weight data corresponding to the activation data cube~(25), updating of a coordinate~(26), completion of data cube computation~(27), permission for next computation~(28).
     & 2\\
    \hline
    CONV\_7 & 
    Initiation of processing~(29), retrieval of specific weight data~(30), verification of specific weight data retrieval~(31).
     & 2,3\\
    \hline
    CONV\_8 & 
    Setting of output channel number to WT FSM~(32), initiation of processing~(33), state change for weight data acquisition~(34), acquisition of the final weight data~(35), completion of weight data acquisition~(36), 
    conclusion of all computations~(37).
     & 2,3\\
    \hline
    CONV\_9 & 
    Initiation of processing~(38), output of specific data~(39), completion of specific data output~(40).
     & 2,3\\
    \hline
    CONV\_10 & 
    Initiation of processing~(41), permission for processing from Delay MAC.~(42), initiation of primary output~(43).
     & 2,3\\
    \hline
    CONV\_11 & 
    Initiation of processing~(44), output of specific data from Delay MAC~(45), primary output of specific data~(46).
     & 2,3\\
    \hline
    CONV\_12 & 
    Retrieval of the final activation data~(47), output of specific data~(48), completion of specific data output~(49).
     & 2,4\\
    \hline
    CONV\_13 & 
    Initiation of processing~(50), retrieval of specific data cube~(51), updating of weight data~(52).
     & 2,3\\
    \hline
    CONV\_14 & 
    Initiation of processing~(53), completion of weight data retrieval~(54), updating of specific address for activation data~(55).
     & 2,3\\
    \hline
  \end{tabularx}  
}
\end{table*}

\subsubsection{State-sequences}
As explained in Section~\ref{sec_acq_normal_state_seq}, we define a new state by combining selected signals in a three-level hierarchy. First, we focus on the primary outputs: nine control signals in the convolutional layer module connected to the “Control module” and “Direct Memory Access module.” These signals include module state, communication (ready, valid), and data address bits. Combined, they form 79 bits.

From an RTL simulation of a $32\times32\times1$ activation input, we obtain 322 normal state sequences using Algorithm~\ref{alg_normal_seq}. An error detector using all 79 bits consumes 29.3\% of the circuit area, which is often impractical. Hence, we reduce overhead by selecting one bit from each of the nine outputs, creating a 9-bit state. Similarly, we choose one bit from other monitored signals. We typically pick either the most or least significant bit or restrict our choice to the bit range in use, then take the top or bottom bit in that range. Similar analyses are conducted for the remaining hierarchies.

Table~\ref{tab_state_seq_conv} lists the monitored signal candidates at three hierarchical levels, along with their normal state sequences and area overhead. $\mathit{State\ bit}$
 indicates the number of bits per state, $\SEQUENCES$ is the count of normal state sequences from the golden RTL simulation, and $\AREA$ shows the fraction of circuit area used by the error detector. At hierarchical Level~1~(1), all bits of primary outputs are used, while Level~1~(2) considers only their most significant bits. At Level~2, states are defined by sub-module outputs (WT FSM, Data FSM, Delay CTR, MAC CTR, and Accumulation). At Level~3, control registers are targeted, including only MSBs (2), only MSBs within the utilized bit range (3), or least significant bits (4).

The control registers, including those for activation and weight data addresses, exhibit periodic value changes during convolution. As a result, LSBs often change more frequently, while MSBs change less. However, Table~\ref{tab_state_seq_conv} shows that monitoring MSBs yields more normal state sequences, causing higher area overhead. Area overhead does not scale linearly with these sequences: at Level~3~(3), the sequence count is eight times that of Level~3~(2), but overhead reaches 20 times. This largely stems from Vivado's optimization. Higher-level hierarchies typically use fewer bits per state and produce fewer normal sequences, thus reducing overhead. Bit selection also matters: for example, Level~1~(1) uses more area than Level~1~(2), so other test circuits select bits to curb overhead. As shown in Section~\ref{sec_fi_result_seq_conv}, there is no major difference among types 2, 3, and 4, making type 2 the primary choice.

\begin{table}
\small
  \caption{Normal state sequences in Conv.}
\label{tab_state_seq_conv}
  \begin{tabularx}{\linewidth}{lXccc} 
    \hline
     \makecell[l]{Hierarchy\\(bit type)} & Target signals & State bits & $\SEQUENCES$ & $\AREA$~(\%)\\
    \hline
    Level 1 (1) & \makecell[l]{Primary outputs} & 79 & 322 & 29.3\\
    \hline
    Level 1 (2) & \makecell[l]{Primary outputs\\(only MSBs)} & 9 & 19 & 0.1\\
    \hline
    Level 2 (2) & \makecell[l]{Outputs in sub-\\modules\\(only MSBs)} & 21 & 76 & 0.3\\
    \hline
    Level 2 (3) & \makecell[l]{Outputs in sub-\\modules\\(only MSBs in the\\used bit range)} & 21 & 293 & 0.4\\
    \hline
    Level 3~(2) & \makecell[l]{Control registers\\(only MSBs)} & 29 & 112 & 0.3\\
    \hline
    Level 3~(3) & \makecell[l]{Control registers\\(only MSBs in the\\used bit range)} & 29 & 897 & 6.4\\
    \hline
    Level 3~(4) & \makecell[l]{Control registers\\(only LSBs)} & 29 & 445 & 3.7\\
    \hline
  \end{tabularx}
  \vspace*{-0mm}
\end{table}

\subsection{Gaussian blur}
\begin{figure}
  \centering
  \includegraphics[width=.9\linewidth]{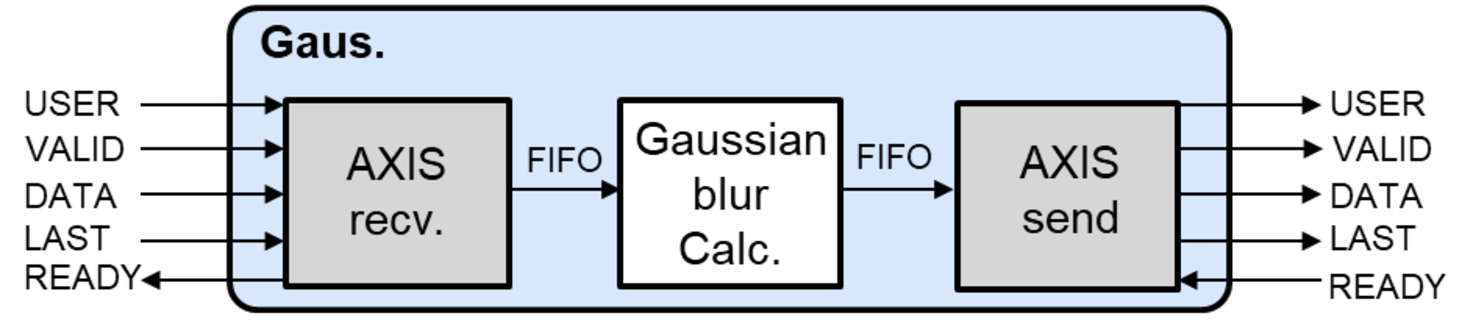}
  \vspace*{-0mm}
  \caption{Architecture of Gaussian blur. A white box is the main module.}
  \label{fig_gaus}
\end{figure}

To evaluate the applicability to general image processing, we focus on Gaussian blur filtering. The Gaussian blur (Gaus.) architecture is shown in Fig.\ref{fig_gaus}. Gaus. processes input image data using the AXI4-Stream protocol, which includes data transmission and control signals (user, valid, last, ready). It receives 
$64\times48$ input images. The AXIS recv. module acquires input via AXI4-Stream and passes it to the Gaussian blur calc. module through FIFOs. The AXIS send module then outputs the results via AXI4-Stream. This pipeline achieves per-clock, per-pixel processing\cite{Yamawaki2018}.

The Gaussian blur module is a specialized module for filter operations, and thus the input patterns it accepts are highly limited, with control flow being entirely independent of the data. Therefore, we define the input patterns, when connected with a test pattern generator as the upstream circuit, as the only valid input patterns

\subsubsection{Petri-nets}
Table~\ref{tab_construction_pn_gaus}  lists the monitored event sets, their corresponding IDs, and the event assignment types used. Three Petri nets are generated, each associated with an ID. These three modules operate synchronously for pipeline processing. By including signals from each module in the event set, their synchronized operation is effectively monitored.
In logic synthesis, 581 LUTs and 577 FFs for Gaus., and 493 LUTs and 329 FFs for 3 Petri nets are utilized in the Zynq-7000 FPGA with part xc7z020clg484-1.
The FPGA differs in convolutional layer computation and Gaussian blur, it essentially does not affect the error detection performance of Petri nets.

\begin{table}
\small
  \caption{Monitored event sets in Gaussian blur.}
  \label{tab_construction_pn_gaus}
\resizebox{\linewidth}{!}{
  \begin{tabularx}{\linewidth}{lXl}  
    \hline
    ID & Event set with Event Label~(\#) & Type\\
    \hline
    GAUS\_1 & 
    Targeting per-line processing, initiation in AXIS recv.~(1), initiation in Gaussian blur calc.~(2), initiation in AXIS send~(3), completion in AXIS send~(4).
    & 2\\
    \hline
    GAUS\_2 & 
    Targeting vertical counters, update in AXIS recv.~(5), update in Gaussian blur calc.~(6), update in AXIS send~(7). Acquisition of specific pixel data in AXIS recv.~(8).
    & 1,3\\
    \hline
    GAUS\_3 & 
    Targeting specific pixel data, acquisition in AXIS recv.~(9), writing to FIFO in AXIS recv.~(10), acquisition in Gaussian blur calc.~(11), writing to FIFO in Gaussian blur calc.~(12), acquisition in AXIS send~(13), output in AXIS send~(14), completion of per-line processing in AXIS send~(15). Completion of image processing~(16).
    & 2,3\\
    \hline
  \end{tabularx}  
}
\end{table}

\subsubsection{State-sequences}
Table~\ref{tab_state_seq_gaus} presents the normal state sequences and corresponding area overheads based on monitored signals from three hierarchical levels, following the same structure as Table~\ref{tab_state_seq_conv}. At Hierarchical Level~1, the primary outputs include four AXI4 stream signals (user, ready, valid, and last). Due to the small number of state bits and limited state sequences, the area overhead is minimal. At Level~2, 21 output signals from sub-modules (AXIS recv., Gaussian blur calc., AXIS send) are monitored, resulting in the most normal sequences and the highest area overhead. In contrast, Level~3 focuses on 11 control registers in the Gaussian blur calculation module, yielding relatively lower overhead.

\begin{table}
\small
  \caption{Normal state sequences in Gaus.}
\label{tab_state_seq_gaus}
  \begin{tabularx}{\linewidth}{lXccc} 
    \hline
    \makecell[l]{Hierarchy\\(bit type)} & Target signals & State bits & $\SEQUENCES$ & $\AREA$~(\%)\\
    \hline
    Level 1 (1)& Primary outputs & 4 & 33 & 0.9\\
    \hline
    Level 2 (1)& \makecell[l]{Outputs in\\sub-modules} & 21 & 57 & 8.5\\
    \hline
    Level 3 (2)& \makecell[l]{Control registers\\(only MSBs)} & 11 & 47 & 4.3\\
    \hline
  \end{tabularx}
  \vspace*{-0mm}
\end{table}

\subsection{Advanced encryption standard~(AES)}
\begin{figure}
  \centering
  \includegraphics[width=.7\linewidth]{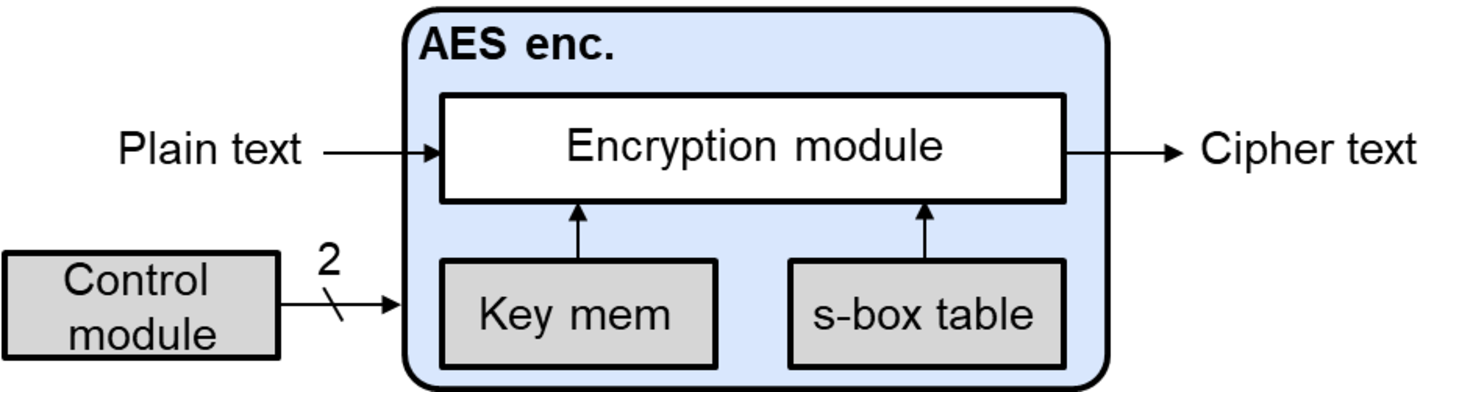}
  \caption{Architecture of AES encryption. A white box is the main module.}
  \label{fig_aes}
\end{figure}

As a de facto security primitive, we target Advanced Encryption Standard (AES) encryption.
The architecture of the AES encryption system implemented is depicted in Fig.~\ref{fig_aes}, as described in \cite{Degnan2021}. 
Our focus was to construct a Petri net that targets the sequential encryption of five 128-bit plaintexts.
The input patterns for the AES encryption module are assumed to be continuous plaintext inputs, synchronized with the state changes that the AES encryption module can accept.
In other words, interruptions during data input are not considered.

\subsubsection{Petri-nets}
Table~\ref{tab_construction_pn_aes} details the monitored event sets. Using the Vivado tool, the AES encryption system and its seven associated Petri nets were synthesized, targeting the Zynq-7000 FPGA.
2,525 LUTs and 2,331 FFs for AES enc., and 281 LUTs and 205 FFs for 7 Petri nets are utilized.

\begin{table}
\small
  \caption{Monitored event sets in AES encryption.}
\label{tab_construction_pn_aes}
\resizebox{\linewidth}{!}{
  \begin{tabularx}{\linewidth}{lXl} 
    \hline
     ID & Event set with Event Label~(\#) & Type\\
    \hline
    AES\_1 & 
    Initiation of processing~(1), permission for state 
    change~(2), permission for round update~(3), update of 
    round~(4).
     & 2\\
    \hline
    AES\_2 & 
    Initiation of processing~(5), permission for state
    change~(6), reset of per-round S-box~(7).
     & 2,3\\
    \hline
    AES\_3 & 
    Initiation of processing~(8), permission for state 
    change~(9), acquisition of per-round plaintext~(10).
     & 2,3\\
    \hline
    AES\_4 & 
    Initiation of processing~(11), start of processing per-plaintext~(12), permission for next plaintext~(13),
    completion per-plaintext~(14).
     & 2,3\\
    \hline
    AES\_5 & 
    Initiation of processing~(15), increment of per-round S-box counter~(16), acquisition of S-box for next round~(17).
     & 2,3\\
    \hline
    AES\_6 & 
    Initiation of processing~(18), permission to update round counter~(19), update of round counter~(20).
     & 2,3\\
    \hline
    AES\_7 & 
    Permission for state change~(21), state change for processing~(22).
     & 2,3\\
    \hline
  \end{tabularx}
}
\end{table}

\subsubsection{State-sequences}

Table~\ref{tab_state_seq_aes} presents the normal state sequences and corresponding area overheads using monitored signals at each hierarchical level, following the same structure as Table~\ref{tab_state_seq_conv}. Since the AES encryption circuit primarily involves simple data passing, only Hierarchical Levels~2 and 3 are considered. Level~2 monitors outputs from sub-modules (Encryption module, Key-mem, and S-box table), where the number of state bits is small, resulting in low area overhead. In Level~3~(1), all control register bits are used, but the increase in normal sequences is limited, keeping area overhead minimal. Level~3~(2) focuses on the most significant bits of control registers in the Encryption module. With only four registers, the area overhead remains very small.

\begin{table}
\small
  \caption{Normal state sequences in AES.}
\label{tab_state_seq_aes}
  \begin{tabularx}{\linewidth}{lXccc} 
    \hline
     \makecell[l]{Hierarchy\\(bit type)} & Target signals & State bits & $\SEQUENCES$ & $\AREA$~(\%)\\
    \hline
    Level 2~(1)& \makecell[l]{Outputs in\\sub-modules} & 3 & 12 & 0.1\\
    \hline
    Level 3~(1) & \makecell[l]{Control registers\\(all bits)} & 9 & 55 & 1.0\\
    \hline
    Level 3~(2) & \makecell[l]{Control registers\\(only MSBs)} & 4 & 15 & 0.2\\
    \hline
  \end{tabularx}
  \vspace*{-0mm}
\end{table}

\subsection{Network-on-Chip (NoC) router}
\begin{figure}
  \centering
  \includegraphics[width=\linewidth]{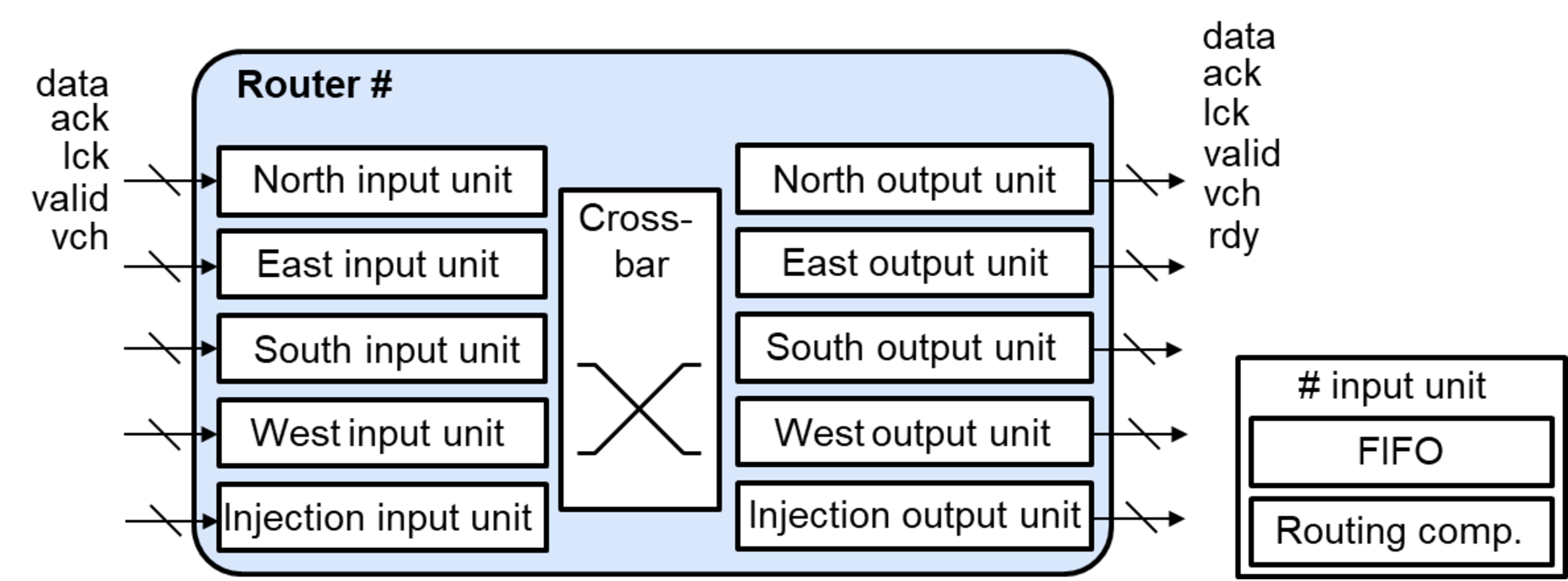}
  \vspace*{-0mm}
  \caption{NoC router architecture.}
  \label{fig_router}
  \vspace*{-0mm}
\end{figure}

The final target is a Network-on-Chip (NoC) router~\cite{Kyle}, an open-source on-chip router originally developed in~\cite{Matsutani2009}. Its architecture is shown in Fig.~\ref{fig_router}. The router connects in five directions—north, east, south, west, and injection—in a mesh-structured NoC. The directional ports connect to neighboring routers, while the injection port links to a directly interfacing module.
Each port has identical input/output signals: "data" for transferred data, "ack" for acknowledgment, "lck" for lock status, "valid" for valid data, and "vch" for the virtual channel, with two virtual channels per port. All ports include an input unit with a FIFO and a routing computation module. YX routing with static arbitration is used, prioritizing multicast over unicast. The routing priority is injection, west, south, east, and north.
The crossbar module, a multiplexer-based controller, handles data transmission. Each port also includes an output unit managing FIFO data ejection and virtual channel switching. Output signals mirror input meanings, with "rdy" indicating the router's ready status.

\begin{figure}
  \centering
  \includegraphics[width=.9\linewidth]{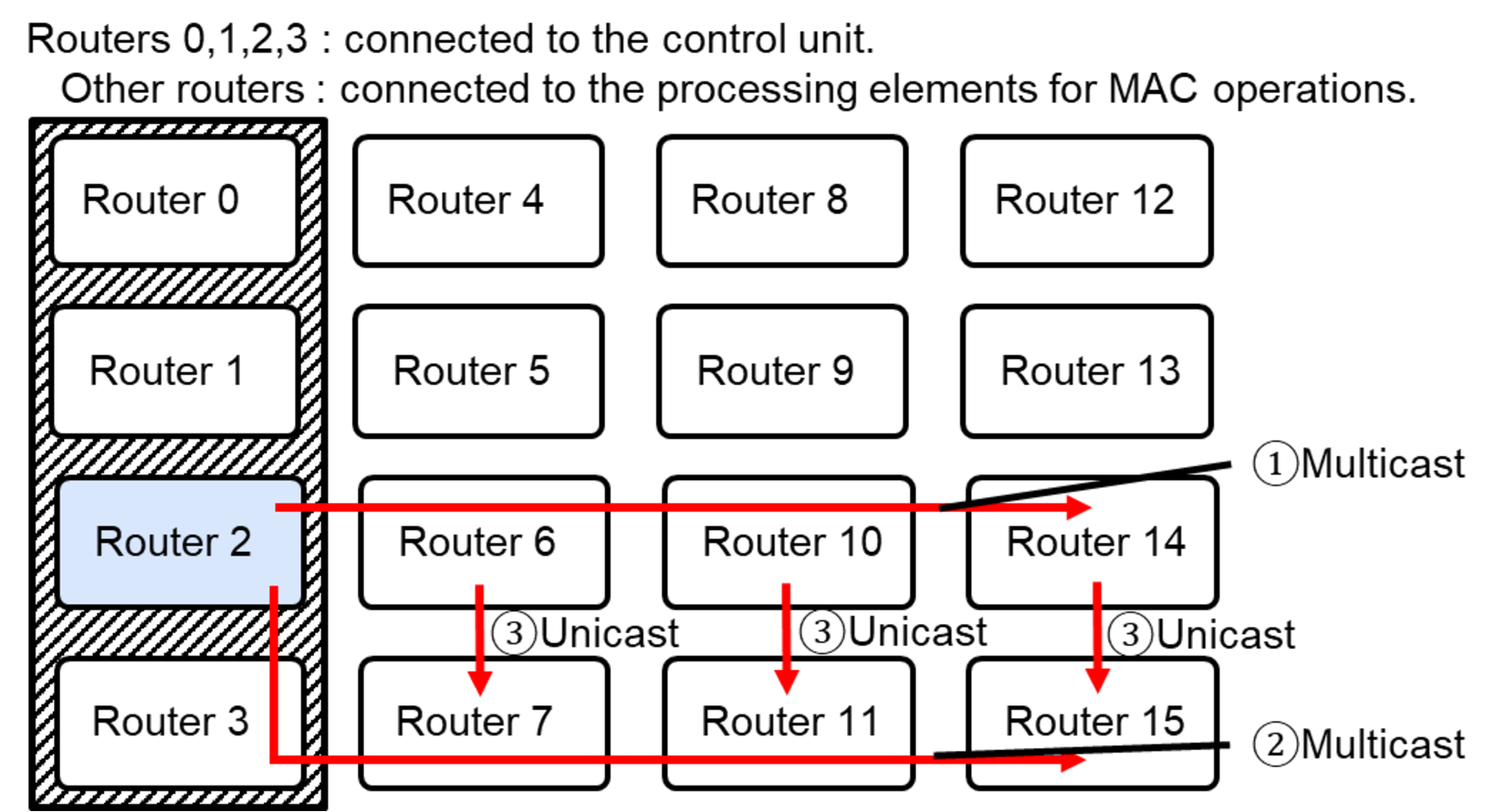}
  \vspace*{-0mm}
  \caption{NoC architecture and simulation scenario.}
  \label{fig_network_on_chip}
  \vspace*{-0mm}
\end{figure}

To develop an error detector for specific applications, we assume the NoC-based CNN accelerator configuration from~\cite{Shao2019}. Fig.~\ref{fig_network_on_chip} shows the assumed setup and data transmission pattern. The NoC is a 4~$\times$~4 mesh of routers. The leftmost column connects to control units (RISC-V processor and global PE), while the others link to MAC processing elements.

Router~2 is monitored using Petri nets and normal state transitions, based on the scenario below.
The simulation models CNN data transmission as in~\cite{Kyle}. Router~2 performs multicast to Routers 6, 10, 14, 7, 11, and 15, representing activation data delivery to MAC units. After a fixed delay, unicast transmissions occur from Router~6 to 7, 10 to 11, and 14 to 15, simulating result transfers between MAC elements. The delay between unicast starts reflects MAC operation latency.
Router~2 sends 128 packets (9 flits each, 66 bits per flit: 2-bit type + 64-bit data) via multicast. Unicast sends 256 flits. A fixed delay of 1,024 cycles models the 128 MAC operations\cite{Shao2019}. Two virtual channels, alternately assigned to each flit, improve throughput and reduce congestion.

\subsubsection{Petri-nets}\label{sec_pn_router}
Table~\ref{tab_construction_pn_router} summarizes 12 monitored event sets. R\_9 and R\_10 each map to two virtual channels, resulting in 14 Petri nets used to monitor all directional ports of Router~2 in the defined scenario.
The test patterns mainly involve multicast data reception and transmission, along with some 9-flit unicast transfers. These tests do not aim to cover all possible communication patterns found in the router specification document but focus on those likely to occur under the assumed configuration.
%
Due to static arbitration, routers in the leftmost column are always prioritized. All 14 Petri nets were successfully validated, confirming their effectiveness for the specified test case.

Router~2 and its 14 associated Petri nets, each corresponding to a directional port, were synthesized using the Vivado tool, targeting xc7k70tfbv676-1. The synthesis results indicate that 1,121 LUTs and 855 FFs were utilized for Router~2, while 415 LUTs and 545 FFs were allocated for the Petri nets.

\begin{table}
\small
  \caption{Monitored event sets in NoC router.}
\label{tab_construction_pn_router}
\resizebox{\linewidth}{!}{
  \begin{tabularx}{\linewidth}{lXl} 
    \hline
     ID & Event set with Event Label~(\#) & Type\\
    \hline
    R\_1 & 
    Permission for transmission~(1), transition to transmission-enabled state~(2).
     & 2\\
    \hline
    R\_2 & 
    Output of final flit (3), transition to transmission-disabled state (4).
     & 2\\
    \hline
    R\_3 & 
    Assertion of enable signal~(5),	update of multicast management signal~(6).
     & 1,2\\
    \hline
    R\_4 & 
    Enabling  of FIFO read~(7), emptiness of FIFO~(8).
     & 2\\
    \hline
    R\_5 & 
    Enabling  of FIFO write~(9), update of FIFO address~(10).
     & 1,2\\
    \hline
    R\_6 & 
    Change of permitted port~(11), storage of previously used port~(12).
     & 1\\
    \hline
    R\_7 & 
    Enabling  of input~(13), Enabling  of output~(14).
     & 1\\
    \hline
    R\_8 & 
    Transmission of tail flit~(15), change of virtual channel~(16).
     & 1,2\\
    \hline
    R\_9 & 
    Transmission of flit~(17), update of flit counter~(18).
     & 2\\
    \hline
    R\_10 & 
    Reception of flit~(19), update of flit counter~(20).
     & 2\\
    \hline
    R\_11 & 
    Change of acknowledgment signal~(21), enabling  of output~(22).
     & 1\\
    \hline
    R\_12 & 
    Change of acknowledgment signal~(23), change of virtual channel~(24).
     & 1\\
    \hline
  \end{tabularx}
}
\end{table}

\subsubsection{State-sequences}
Table~\ref{tab_state_seq_router} shows the normal state sequence and its area overhead. 
The sequence is constructed for each directional port, focusing solely on primary output signals. Including internal signals—such as those from submodules or control registers—would make it difficult to pass the test patterns described in Section~\ref{sec_pn_router}, so they are excluded. The normal state sequence using only primary outputs is validated with the same test patterns in Section~\ref{sec_pn_router}.
The area overhead exceeds 10\%, primarily because the sequence must capture diverse behaviors, including both multicast and unicast data transfers.

\begin{table}
\small
  \caption{Normal state sequences in NoC router.}
\label{tab_state_seq_router}
  \begin{tabularx}{\linewidth}{lXccc} 
    \hline
     \makecell[l]{Hierarchy\\(bit type)} & Target signals & State bits & $\SEQUENCES$ & $\AREA$~(\%)\\
    \hline
    Level 1 (1)& Primary outputs & 8 & 18 & 12.5\\
    \hline
  \end{tabularx}
\end{table}

\section{Experimental results}\label{sec_experimental_results}
\subsection{Experimental setup} \label{sec_experimental_setup}
To assess the efficacy of our error detectors, we perform RTL fault injection simulations on the four target designs in two cases. 

In Case~1, faults are injected into control registers within the target design, assuming a direct impact of soft errors on the target design.
For experimental efficiency, we limited fault injections to the main processing module of each design. 

In Case~2, faults are injected into the primary control inputs of the target design for Conv., Gaus., and AES enc., assuming that faults are propagating from upstream circuits. We intentionally randomized primary control inputs across ten consecutive cycles. To justify this fault injection approach, we conducted preliminary experiments involving over 600 instances of bit-flip fault injections on parts of the control modules of Conv. and AES enc., resulting in erroneous primary control inputs appearing for more than 14,512 and 15 cycles on average, respectively. Similarly, the bit-flip in Gaus. exhibited prolonged incorrect outputs on AXI4 streams, implying that primary control inputs can receive similar faults. These results indicate that our fault injection setup in Case~2 is not excessive but practical.
Unlike the other three circuits, the control input of Router~2 is not derived solely from a single upstream circuit, but is connected to adjacent routers. To more accurately assess the impact of failures in the surrounding routers, faults were injected into two neighboring routers (Router~3 and Router~6), and the resulting faults that appeared on the control input of the target router (Router~2) were treated as test input faults. In other words, four input signals from each of the two directional ports in Router~2, which are connected to Router~3 and Router~6, were targeted.

\subsection{Error detection performance of Petri nets}
\label{sec_peformance_pn}
\subsubsection{All Petri nets}
For Case 1 fault injection, Table~\ref{tab_fi_control_regs} presents the error detection performance across the four designs, using all Petri nets. $\NREGS$($\mathit{N_{bits}}$) denotes the number of control registers and their total bit count targeted for fault injection.  
43,500 faults were injected for the Conv., 72,600 for the Gaus., 40,000 for the AES enc., and 31,800 for the Router, evenly distributed across target registers within each design.
$\NOE$ indicates the number of output errors, including both incorrect computation results and abnormal terminations.
Among 
$\NOE$, 87.6\% of errors in Conv., 68.6\% in Gaus., and 81.4\% in AES enc. were due to incorrect results. 
The remaining cases reflect abnormal termination. In the Router, 85.7\% of $\NOE$ involve incorrect data transfers—such as data corruption or missing flits—while 14.3\% are due to processing timeouts.
$\DR$ denotes the error detection rate. 
$\DRTO$, a subset of $\DR$, indicates cases where Petri nets detect only the final incorrect transition. Since the timing of this check varies by case, a high $\DRTO$ often implies longer error detection latency. The latency, $\LATENCY$, is calculated excluding detections counted in $\DRTO$.

Regarding $\DR$, Conv. achieved the highest detection rate at 99.5\%, attributed to its larger number of Petri nets. In contrast, Gaus. had the lowest $\DR$ at 88.0\%, possibly due to its limited diversity across only three Petri nets.
Given the clock cycles for normal operation—20,521 for Conv., 8,676 for Gaus., 432 for AES enc., and 2,798 for Router—the $\LATENCY$ remains low across all designs, indicating fast error detection. 
$\DRTO$ for Conv. and Gaus. is nearly negligible.
$\DRTO$ is not applicable to the Router, as final-transition-based timeout detection is not implemented due to challenges in uniquely defining the last transition. Nevertheless, the Router achieves over 95\% $\DR$, demonstrating both high accuracy and quick detection.

Table~\ref{tab_fi_primary_in} presents the error detection performance for Case~2. 
$\NINPUT$ indicates the number of targeted primary control inputs. Fault injections totaled 10,000 for Conv., 10,000 for Gaus., 40,000 for AES enc., and 30,000 for the Router.
In $\NOE$, 100.0\% of errors in Conv., 98.6\% in Gaus., and 92.6\% in AES enc. were incorrect results, with the rest due to abnormal termination. For the Router, 88.8\% of $\NOE$ involved incorrect data transfers (e.g., corruption or missing flits), while the remainder were processing timeouts.

The $\DR$ exceeded 96\% in Conv., Gaus., and AES enc. With short $\LATENCY$, $\DRTO$ was 0.0\% for Conv. and AES enc., and 0.3\% for Gaus., indicating fast and effective error detection.
However, the Router’s 
$\DR$ was lower, likely due to its more complex control flow. Its internal state machine supports various transmission scenarios, making abnormal behavior harder to capture with Petri nets. Limiting communication patterns and monitoring more specific behaviors could improve detection performance.

Note that redundancy techniques like simple TMR cannot mitigate these faults, as all modules receive the same faulty inputs and fail identically. In contrast, Petri nets detect control-flow disturbances caused by faulty inputs and the resulting incorrect outputs. They are effective not only for Conv., Gaus., and AES enc., which show high detection rates, but also for routers, despite their relatively lower detection rate.

It should be noted that although Petri net detectors may be affected by soft errors and produce false negatives, they do not interfere with the monitored circuit, as they have no outputs that feed back into it.

\begin{table}
\small
  \caption{Petri-net-based error detection performance in Case~1: detection rate $\DR$, detection rate by final incorrect transition $\DRTO$, and error detection latency $\LATENCY$.
  }
 \label{tab_fi_control_regs}
 \resizebox{\columnwidth}{!}{%
  \begin{tabular}{cccccc}
    \hline
    Design & $\NREGS$ ($\mathit{N_{bits}}$) & $\NOE$ & \makecell[c]{$\DR$ \\(\%)} & \makecell[c]{$\DRTO$\\(\%)} & \makecell[c]{$\LATENCY$\\(cycles)} \\
    \hline
    Conv. & 29 (246) &  31,898 & 99.5 & 0.1 &107.6 \\
    \hline
    Gaus. & 11 (35) &  53,638 & 88.0 & 0.5 &53.4 \\
    \hline
    AES enc.& 4 (9) &  26,240 & 95.3 & 12.0 &3.8 \\
    \hline
    Router& 106 (321) & 4,288 & 95.4 & N/A &7.8 \\
    \hline
  \end{tabular}
 }
\end{table}

\begin{table}
\centering
\small
  \caption{Petri-net-based error detection performance in Case~2.
  }
 \label{tab_fi_primary_in}
  \begin{tabular}{cccccc}
    \hline
     Design & $\NINPUT$ & $\NOE$ & \makecell[c]{$\DR$ \\(\%)} & 
 \makecell[c]{$\DRTO$\\(\%)} & \makecell[c]{$\LATENCY$\\(cycles)}\\
    \hline
    Conv. & 8  & 9,911 & 99.9 & 0.0 &2.5\\
    \hline
    Gaus. & 4  & 9,668 & 96.3 & 0.3 &102.1\\
    \hline
    AES enc.& 2  & 18,310 & 99.9 & 0.0 &1.0\\
    \hline
    Router & 4~$\times$~2 & 4,538 & 47.8 & N/A &152.9 \\
    \hline
  \end{tabular}
\end{table}

\subsubsection{Trade-off between detection rate and area overhead}
\begin{figure}
  \centering
  \includegraphics[width=\linewidth]{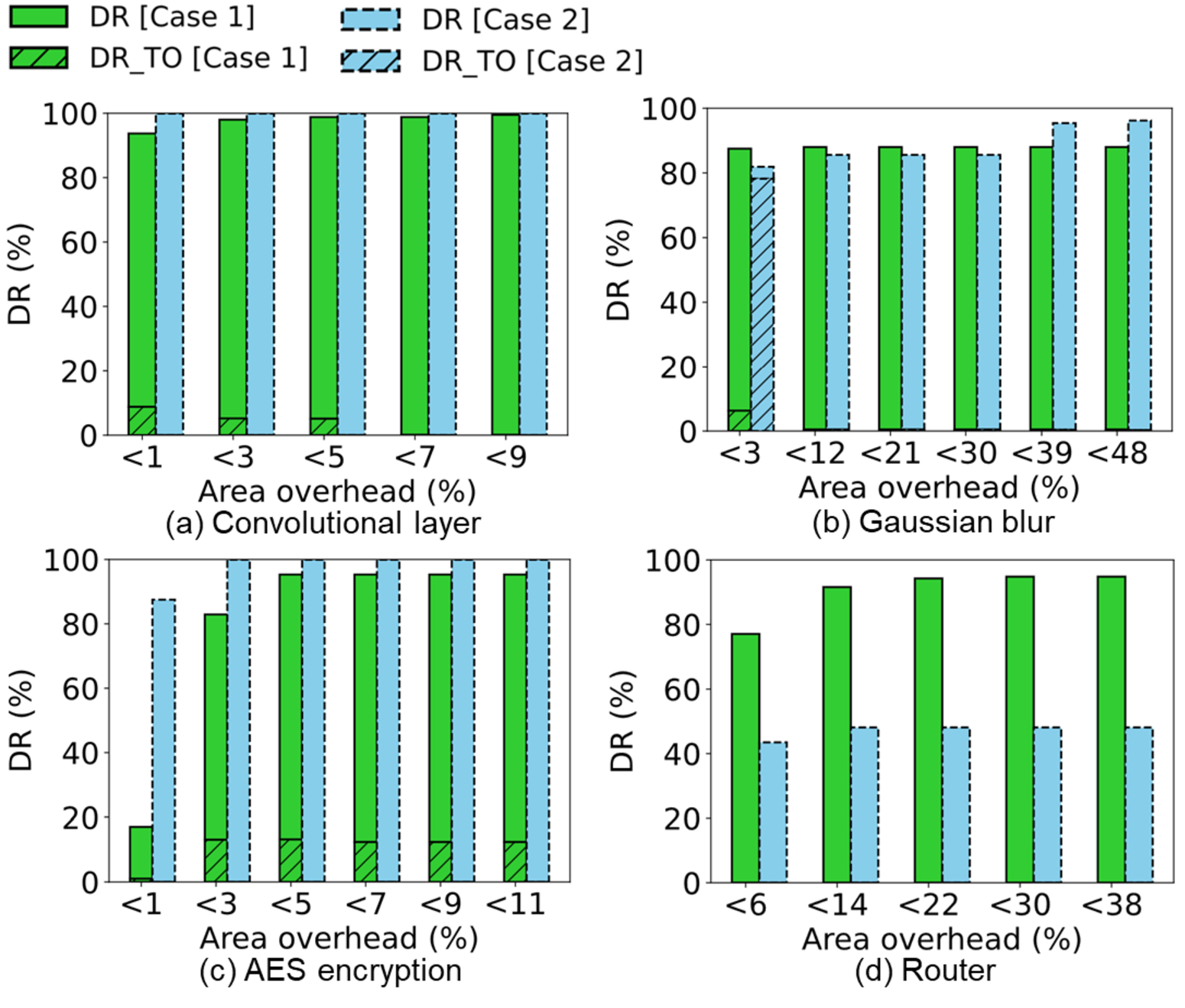}
  \caption{Trade-off between area overhead and DR in Petri-net-based method.}
  \label{fig_tradeoff_area_DR}
  \vspace*{-0mm}
\end{figure}

We next selectively choose Petri nets to balance area overhead and error detection.
Fig.~\ref{fig_tradeoff_area_DR}(a) shows the relationship between area overhead and $\DR$ for convolutional layer computation. Area overhead is based on LUT count.
The x-axis indicates area overhead thresholds. The left bars show the maximum $\DR$ and subset $\DRTO$ for each threshold in Case 1; the right bars show the same for Case~2.
The maximum $\DR$ at each threshold is determined by evaluating all Petri net combinations within the given area limit. Using all 14 Petri nets results in 9\% area overhead. In Case 1, $\DR$ increases with area up to 9\%, while $\DRTO$ decreases, indicating efficient improvement.
Remarkably, just 1\% area achieves 93.7\% 
$\DR$ in Case 1 and 99.9\% in Case 2. With 
$\DRTO$ consistently at 0\% in Case~2, errors are detected rapidly.

The relationship between area overhead and $\DR$ for Gaussian blur is shown in Fig.~\ref{fig_tradeoff_area_DR}(b).
In Case 1, the maximum $\DR$ is achieved with up to 12\% area overhead, with no further improvement beyond that point. In both Case 1 and Case 2, $\DRTO$ drops significantly above 12\% overhead, indicating improved error detection latency. In Case 2, $\DR$ exceeds 80\% with less than 3\% overhead and continues to improve gradually.

Fig.~\ref{fig_tradeoff_area_DR}(c) shows the results for AES encryption.
Using all seven Petri nets results in 10\% area overhead. However, in Case 1, the maximum $\DR$ is achieved with less than 5\% overhead, while $\DRTO$ remains around 13\%.
Notably, $\DR$ increases sharply with each 1\% area increment up to 3\%.
In Case~2, the maximum $\DR$ reaches 99.9\% at under 3\% overhead, with $\DRTO$ consistently at 0\%, indicating rapid error detection.

Fig.\ref{fig_tradeoff_area_DR}(d) shows the result of Router.
The same Petri nets are implemented for each direction of the router, as similar communication may occur at each port.
The area overhead when using all seven Petri nets is approximately 37\%. In Case~1 the maximum $\DR$ is reached when the area overhead is $<$30\%.
While $\DR$ for Case~2 is relatively low, maximum $\DR$ is reached at $<$14\%.

These results demonstrate that the proposed method allows for effective consideration of adding or removing Petri nets based on the trade-off between area overhead and error detection rate.
This enables flexible adaptation to circuit area constraints and error detection rate requirements.

\subsection{Error detection performance of state-sequences}
\subsubsection{Convolutional layer computation}
\label{sec_fi_result_seq_conv}
Table~\ref{tab_seq_area_DR_conv} shows the error detection performance based on the normal state sequence for the convolutional layer.
37,700 and 10,000 faults were injected for Case 1 and Case 2, resulting in 27,679 and 9,915 failures, respectively.
The leftmost column indicates the hierarchy level and monitored bit type, corresponding to Fig.~\ref{fig_hierarchy} and Table~\ref{tab_monitored_bit_types}. The parameters $\NINJ$, $\NOE$, $\DR$, $\DRTO$, and $\LATENCY$ follow the definitions in Table~\ref{tab_fi_control_regs}, with the same fault injection targets.

In Case 1, when using Level 1 signals (primary outputs), $\DR$ is relatively low—even when all bits are monitored, as shown in Level 1(1). Given this low $\DR$ and the high area overhead from Table~\ref{tab_state_seq_conv}, primary output signals may be unsuitable for practical applications.
Using submodule outputs (Level 2) improves $\DR$, though it stays below 90\%. With control registers (Level 3), $\DR$ exceeds 85\%, peaking in Level 3(3), where MSBs within the used bit range are targeted. The trade-off between $\DR$ and area overhead is further discussed in Section~\ref{sec_comp_pn_normal_seq_DR_area}.

In Case~2, $\DR$ exceeds 90\% across all hierarchy levels and bit selection types.
However, $\DRTO$ remains consistently above 90\%, indicating that normal state sequences contribute little to rapid error detection.
For Case~2, monitoring only the circuit's final state may be a more effective approach.
The final choice of hierarchy level for normal state sequences should consider both detection performance in Case~1 and area overhead. These factors are further evaluated in Section~\ref{sec_comp_pn_normal_seq_DR_area}.

\begin{table}
\centering
\small
  \caption{Normal-state-sequence-based error detection performance for Conv.}
\label{tab_seq_area_DR_conv}
  \begin{tabular}{cccccc} 
    \hline
     & \makecell[l]{Hierarchy\\(bit type)}  & \makecell[c]{$\DR$\\(\%)} & \makecell[c]{$\DRTO$\\(\%)} & \makecell[c]{$\LATENCY$\\(cycles)} \\
    \hline
    \multirow{7}{*}{Case~1} &Level 1 (1)& 46.4 & 39.3 & 326.5 \\
    \cline{2-5}
     &Level 1 (2)& 44.9 & 39.3 & 1255.3\\
    \cline{2-5}
     &Level 2 (2)& 75.1 & 0.1 & 2330.4\\
    \cline{2-5}
     &Level 2 (3)& 82.1 & 0.1 & 1580.6\\
    \cline{2-5}
     &Level 3~(2)& 85.0 & 0.0 & 1858.1\\
    \cline{2-5}
     &Level 3~(3)& 91.8 & 0.0 & 1392.5\\
    \cline{2-5}
     &Level 3~(4)& 89.4 & 0.0 & 1268.6\\
    \hline
    
    \multirow{7}{*}{Case~2} & Level 1 (1)& 96.1 & 94.9 & 2.1\\
    \cline{2-5}
     &Level 1 (2) & 96.1 &  96.1 & 126.1\\
    \cline{2-5}    
     &Level 2 (2) & 100.0 & 99.9 & 4363.5\\
    \cline{2-5}
     &Level 2 (3) & 100.0 & 99.9 & 1587.5\\
    \cline{2-5}
     &Level 3~(2) & 99.9 & 99.9 & 8417.5\\
    \cline{2-5}
     &Level 3~(3) & 100.0 & 99.9 & 4209.8\\
    \cline{2-5}
     &Level 3~(4) &  99.9 & 99.9 & 2806.8\\
    \hline    
  \end{tabular}
\end{table}

\subsubsection{Gaussian blur}
60,500 and 20,000 faults were injected for Case 1 and Case 2, respectively, resulting in 43,586 and 19,357 failures.
Table~\ref{tab_seq_area_DR_gaus} shows the error detection performance for Gaussian blur. Most $\DR$ values exceed 85\%, except for hierarchy Level~3(3) in Case~2, which targets the MSBs of control registers.
Input faults in Case~2 rarely affect state sequences, whereas control register bit upsets in Case~1 are effectively detected.
Regarding area overhead, Table~\ref{tab_state_seq_gaus} shows that targeting primary outputs (Level~1(1)) requires just 0.9\%, while other levels exceed 4\%.
Thus, Level~1(1) offers the best area efficiency for improving error detection. The consistently low 
$\DRTO$ confirms rapid error detection.
\begin{table}
\centering
\small
  \caption{Normal-state-sequence-based error detection performance for Gaus.}
\label{tab_seq_area_DR_gaus}
  \begin{tabular}{ccccc} 
    \hline
     &\makecell[l]{Hierarchy\\(bit type)} & \makecell[c]{$\DR$\\(\%)} & \makecell[c]{$\DRTO$\\(\%)} & \makecell[c]{$\LATENCY$\\(cycles)} \\
    \hline
     \multirow{3}{*}{Case~1} &Level 1 (1)& 87.7 & 0.5 & 335.8 \\
    \cline{2-5}
     &Level 2 (1)& 87.8 & 0.0 & 480.6\\
    \cline{2-5}
     &Level 3 (2)& 89.8 & 4.6 & 461.9\\
    \hline

    \multirow{3}{*}{Case~2} &Level 1 (1) & 87.3 & 0.0 & 1393.6\\
    \cline{2-5}
     &Level 2 (1) & 98.7 & 0.0 & 69.9\\
    \cline{2-5}
     &Level 3 (2) & 1.9 & 1.7 & 124.8\\
    \hline
  \end{tabular}
\end{table}

\subsubsection{AES encryption}
40,000 and 20,000 faults were injected for Case 1 and Case 2, resulting in 26,240 and 9,105 failures, respectively.
In Case~1, the highest 
$\DR$ is achieved by monitoring all bits of the control registers (Level~3(1)). Although monitoring all bits typically increases area overhead, it remains low at just 1.0\%, as shown in Table~\ref{tab_state_seq_gaus}, due to AES encryption using only four control registers.
In Case~2, Level~3(1) also yields the highest $\DR$, with all $\DR$ values exceeding 85\%. While Level~3(1) has the highest overhead, the increase is modest compared to Level~2(1) and Level~3(2).
Thus, monitoring all bits of control registers appears practical for normal state sequences.
\begin{table}[t]
\centering
\small
  \caption{Normal-state-sequence-based error detection performance for AES enc.}
\label{tab_seq_area_DR_AES_case1}
  \begin{tabular}{ccccc} 
    \hline
     &\makecell[l]{Hierarchy\\(bit type)} & \makecell[c]{$\DR$\\(\%)} & \makecell[c]{$\DRTO$\\(\%)} & \makecell[c]{$\LATENCY$\\(cycles)} \\
    \hline
     \multirow{3}{*}{Case~1} &Level 2 (1) & 31.1 & 10.0 & 0.3\\
    \cline{2-5}
     &Level 3~(1) & 100.0 & 0.0 & 0.1\\
    \cline{2-5}
     &Level 3~(2) & 47.5 & 0.0 & 0.2\\
    \hline     
     \multirow{3}{*}{Case~2} &Level 2~(1) & 89.0 & 0.0 & 2.8\\
    \cline{2-5}
     &Level 3~(1) & 95.5 & 0.0 & 2.2\\
    \cline{2-5}
     &Level 3~(2) & 86.1 & 0.0 & 2.0\\
    \hline    
  \end{tabular}
\end{table}

\subsubsection{Router}
31,800 and 30,000 faults were injected for Case 1 and Case 2, respectively, resulting in 4,288 and 4,538 failures.
As shown in Table~\ref{tab_seq_area_DR_router}, while 
$\DR$ exceeds 94\% in Case~1, it drops significantly in Case~2. This trend aligns with the Petri net results presented in Section~\ref{sec_peformance_pn}.

\begin{table}[t]
\centering
\small
  \caption{Normal-state-sequence-based error detection performance for Router.}
\label{tab_seq_area_DR_router}
  \begin{tabular}{ccccc} 
    \hline
      & \makecell[l]{Hierarchy\\(bit type)} & \makecell[c]{$\DR$\\(\%)} & \makecell[c]{$\LATENCY$\\(cycles)} \\
    \hline
     Case~1 & Level 1 (1) & 94.5 & 15.2 \\  
    \hline
     Case~2 & Level 1 (1) & 45.7 & 181.3 \\
    \hline
  \end{tabular}
\end{table}

\subsection{Comparison of Petri-net-based error detection and state-sequence-based error detection}
\label{sec_comp_pn_normal_seq_DR_area}
\begin{figure}
  \centering
  \includegraphics[width=\linewidth]{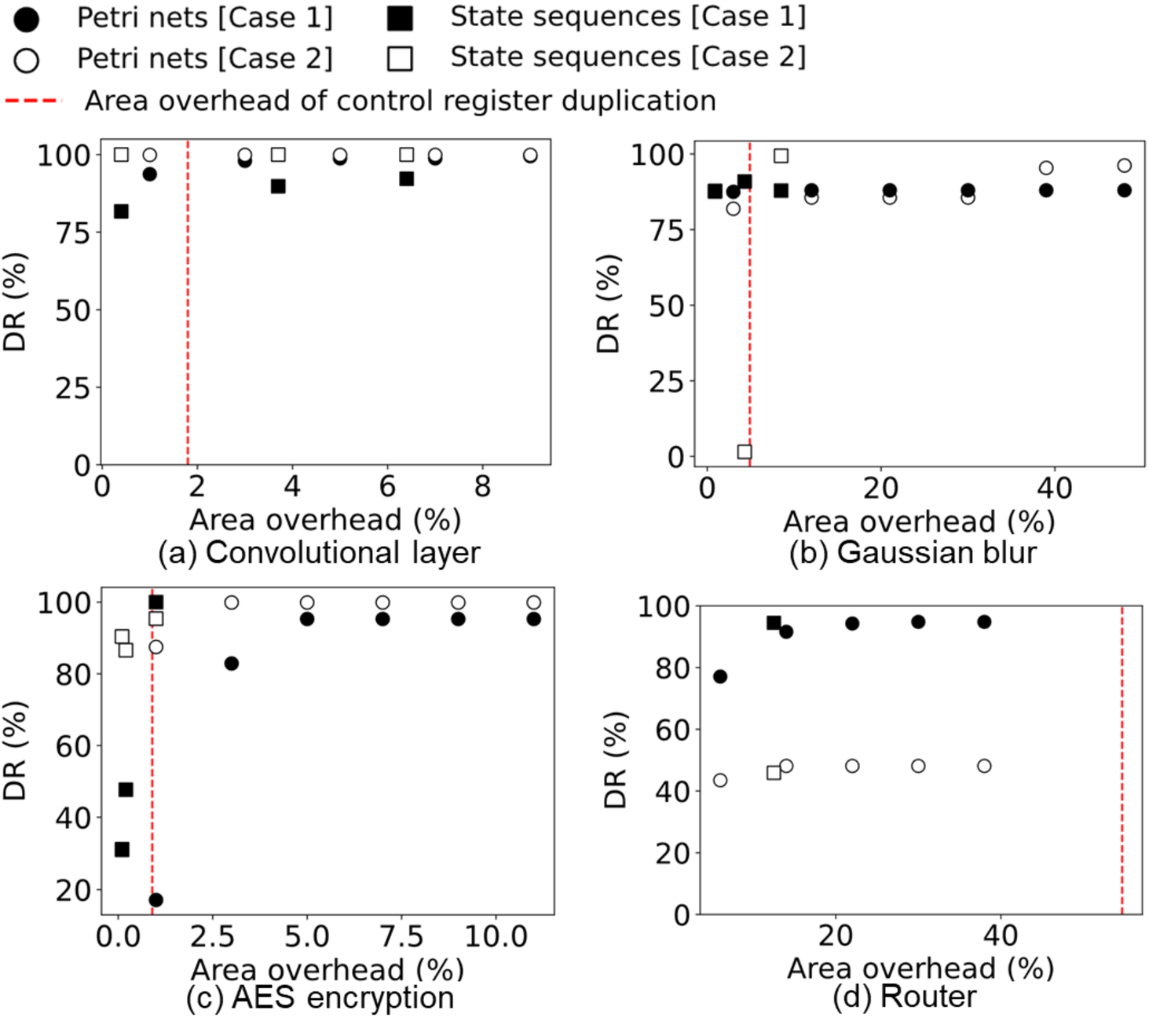}
  \caption{Performance comparison of Petri-nets and state sequences}
  \label{fig_pn_seq_DR_area}
\end{figure}

To evaluate error detection performance, we compare Petri nets and normal state sequences, considering area overhead. As a baseline, we also assess the overhead of duplicating control registers.
Duplicating control registers enables detection of all errors caused by faults in them, achieving a 100\% detection rate in Case~1. However, it fails to detect faults in primary control inputs, resulting in 0\% detection in Case~2.
Fig.~\ref{fig_tradeoff_area_DR} plots the Petri net detection rate against area overhead thresholds.
Fig.~\ref{fig_pn_seq_DR_area} shows detection rates and area overheads for four target designs using both methods. Circles represent Petri nets; rectangles indicate normal state sequences. Black and white show Case~1 and Case~2 results, respectively. Red lines mark the area overhead from duplicating all control registers.

For the convolutional layer, Level~2~(3), Level~3~(3), and Level~3~(4) in normal state sequences are selected for plotting to reduce visual clutter. Duplicating control registers incurs about 2\% area overhead. Below this threshold—left of the red line—both Petri nets and normal state sequences achieve $\DR$ above 80\%, showing effective detection with minimal overhead.
However, in Case~2, $\DR$ from normal state sequences includes a large $\DRTO$ component, leading to longer detection latency. For faster detection, Petri nets may be more effective.
In Case~1, $\DR$ for both methods increases with area overhead, while in Case~2, it remains constant.

Fig.~\ref{fig_pn_seq_DR_area}(b) shows the results for Gaussian blur. Within the range of lower area overheads than control register duplication, normal state sequences achieve higher $\DR$
than Petri nets, with the leftmost point showing the highest $\DR$ for both Case~1 and Case~2.
In Case~1, increasing area does not improve $\DR$, while in Case~2, the highest $\DR$ is reached at 8.5\% overhead using normal state sequences.
As they offer the highest $\DR$ with minimal area cost in both cases, normal state sequences may be more suitable than Petri nets—especially in applications where area is a critical constraint.

Fig.~\ref{fig_pn_seq_DR_area}~(c) presents the results for AES encryption. At an area overhead very close to that of control register duplication, the highest $\DR$ is achieved when using normal state sequences for both Case~1 and Case~2. Increasing the area overhead does not improve the $\DR$ for Case~1, whereas the $\DR$ is improved by using Petri nets in Case~2.

Fig.~\ref{fig_pn_seq_DR_area}(d) shows the results for the NoC Router. The highest $\DR$ for both Case 1 and Case~2 is achieved using normal state sequences at 12.5\% area overhead. Control register duplication incurs significant overhead due to the high number of registers, largely driven by extensive BRAM usage for data storage.
Given this, both Petri nets and normal state sequences offer efficient error detection.
In Case~1, unless area must be minimized, normal state sequences are likely more suitable than Petri nets.
In Case~2, the maximum $\DR$ is lower at 48\%, with Petri nets providing better performance.

The discussion shows that neither the Petri net–based method nor the state sequence–based method is universally superior. The best choice depends on the application's specific constraints and requirements. This work expands the range of error detection options available to designers aiming to enhance chip reliability.

\subsection{Comparison with related work}\label{sec_comparison_with_related_work}
For further evaluations, we compared the proposed methods with a fault detection approach using machine learning–generated assertions~\cite{Pouya2016}. In that work, GoldMine~\cite{Shobha2010} was used to create assertions for three ISCAS benchmark circuits~\cite{Brglez1989}. GoldMine takes Verilog code and Value Change Dump (VCD) files as input. If VCD files are not provided, it can generate them using random RTL simulations.

In our experiments, we used VCD files from both golden and random simulations within GoldMine. Assertions were generated using three mining engines: prism, dtree, and bgdf, with default parameters except for the target cycle, which was set beyond each design’s normal processing cycles.
Besides, the Verilog files for the Convolutional layer and Router were incompatible with GoldMine due to syntax limitations. For the AES encryption circuit, no assertions were generated likely because GoldMine supports only single-bit outputs, and AES has only one single-bit output.
While GoldMine is a powerful tool, it struggles with a wide range of practical circuits and may suffer from excessive runtime, as also noted in~\cite{Pouya2016}.

For the Gaussian blur, GoldMine initially generated 85 assertions. By default, these are validated using a formal verification tool. However, we used golden RTL simulations instead, focusing on detecting control-flow deviations from expected behavior. As a result, 18 assertions were confirmed valid, remaining consistently true during golden simulations.
%
Using the same Case 1 and Case 2 setups described in Section~\ref{sec_experimental_setup}, 17,600 and 10,000 faults were injected, resulting in 12,738 and 9,679 observed errors ($\NOE$), respectively. The 18 previously validated assertions were used for detection, identifying 87.4\% of errors in Case~1 and 82.4\% in Case~2.
To evaluate the trade-off between area overhead and detection rate, we determined the minimum overhead needed to achieve the maximum $\DR$ by testing all combinations of the 18 assertions. This resulted in three assertions selected for Case~1 and five for Case~2.

\begin{figure}
  \centering
  \includegraphics[width=\linewidth]{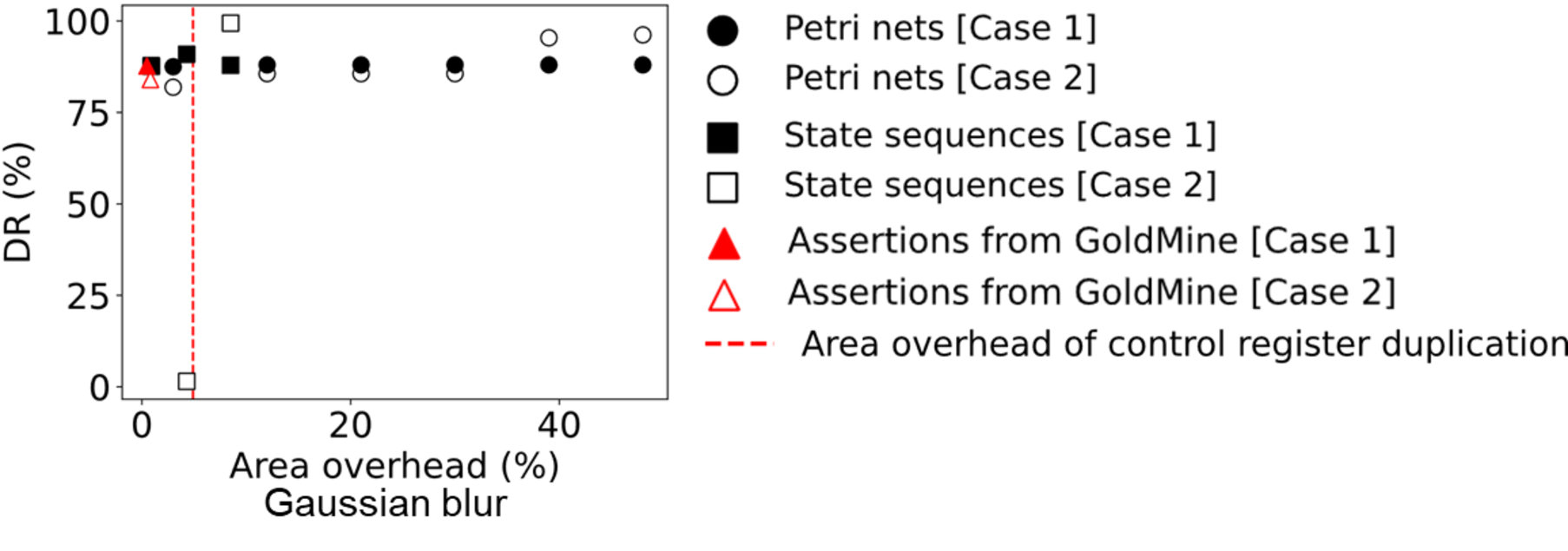}
  \caption{Performance comparison of Petri-nets, state sequences, and assertions from GoldMine~\cite{Shobha2010}}
  \label{fig_pn_seq_goldmine_DR_area}
\end{figure}

Fig.~\ref{fig_pn_seq_goldmine_DR_area} shows the area overheads and detection rates. While the selected assertions have slightly lower overhead than methods like Petri nets or normal state sequences, their detection rates are also lower. When over 90\% detection is required, our proposed methods are more effective. Combining them with assertion-based techniques to boost performance while minimizing overhead is planned as future work.

\section{Conclusion}\label{sec_conclusion}
This paper presented two control-flow-based error detection methods: one using Petri nets generated from specifications and the other using state sequences derived from runtime execution. We developed a methodology for implementing both Petri net detectors and state sequence checkers, validated through fault injection on a convolutional layer, Gaussian blur, AES encryption, and a NoC router. Detection rates ranged from 48\% to 100\% for both register bit-flips and primary input faults, whereas simple register duplication cannot detect errors caused by primary input faults.
Maximum detection was achieved with area overheads of only a few percent to around 10\% in most cases. By selectively applying these methods, designers can explore the reliability-area trade-off.

\bibliographystyle{./bibtex/IEEEtran}
\bibliography{./bibtex/IEEEabrv,mybibfile}

\begin{thebibliography}{10}
\providecommand{\url}[1]{#1}
\csname url@samestyle\endcsname
\providecommand{\newblock}{\relax}
\providecommand{\bibinfo}[2]{#2}
\providecommand{\BIBentrySTDinterwordspacing}{\spaceskip=0pt\relax}
\providecommand{\BIBentryALTinterwordstretchfactor}{4}
\providecommand{\BIBentryALTinterwordspacing}{\spaceskip=\fontdimen2\font plus
\BIBentryALTinterwordstretchfactor\fontdimen3\font minus \fontdimen4\font\relax}
\providecommand{\BIBforeignlanguage}[2]{{%
\expandafter\ifx\csname l@#1\endcsname\relax
\typeout{** WARNING: IEEEtran.bst: No hyphenation pattern has been}%
\typeout{** loaded for the language `#1'. Using the pattern for}%
\typeout{** the default language instead.}%
\else
\language=\csname l@#1\endcsname
\fi
#2}}
\providecommand{\BIBdecl}{\relax}
\BIBdecl

\bibitem{Rech2024}
P.~Rech, ``{Artificial Neural Networks for Space and Safety-Critical Applications: Reliability Issues and Potential Solutions},'' \emph{IEEE Transactions on Nuclear Science}, vol.~71, no.~4, pp. 377--404, 2024.

\bibitem{Jha2019}
{S. Jha, et al.}, ``{ML-Based Fault Injection for Autonomous Vehicles: A Case for Bayesian Fault Injection},'' in \emph{2019 49th Annual IEEE/IFIP International Conference on Dependable Systems and Networks (DSN)}, jun 2019, pp. 112--124.

\bibitem{Fausti2019}
{F. Fausti, et al.}, ``{Single Event Upset tests and failure rate estimation for a front-end ASIC adopted in high-flux-particle therapy applications},'' \emph{Nuclear Instruments and Methods in Physics Research Section A: Accelerators, Spectrometers, Detectors and Associated Equipment}, vol. 918, pp. 54--59, 2019.

\bibitem{Tanaka2021}
{T. Tanaka, et al.}, ``{Impact of Neutron-Induced SEU in FPGA CRAM on Image-Based Lane Tracking for Autonomous Driving: From Bit Upset to SEFI and Erroneous Behavior},'' \emph{IEEE Transactions on Nuclear Science}, vol.~69, no.~1, pp. 35--42, 2022.

\bibitem{Bittel2024}
{B. Bittel, et al.}, ``{Data Center Silent Data Errors: Implications to Artificial Intelligence Workloads \& Mitigations},'' in \emph{2024 IEEE International Reliability Physics Symposium (IRPS)}, 2024, pp. 1--5.

\bibitem{Keller2021}
{A. M. Keller, et al.}, ``{The Impact of Terrestrial Radiation on FPGAs in Data Centers},'' \emph{ACM Trans. Reconfigurable Technol. Syst.}, vol.~15, no.~2, dec 2021.

\bibitem{Konno2024}
{S. Konno, et al.}, ``{Exploration of Fault Identification and Automatic Recovery in Cloud-based FPGA Systems},'' in \emph{2024 IEEE International Conference on Consumer Electronics (ICCE)}, 2024, pp. 1--6.

\bibitem{Hashimoto2020}
M.~Hashimoto and W.~Liao, ``{Soft Error and Its Countermeasures in Terrestrial Environment},'' in \emph{2020 25th Asia and South Pacific Design Automation Conference (ASP-DAC)}, 2020, pp. 617--622.

\bibitem{Muslim2015}
M.~Mustapa and M.~Niamat, ``{Temperature, Voltage, and Aging Effects in Ring Oscillator Physical Unclonable Function},'' in \emph{2015 IEEE 17th International Conference on High Performance Computing and Communications, 2015 IEEE 7th International Symposium on Cyberspace Safety and Security, and 2015 IEEE 12th International Conference on Embedded Software and Systems}, 2015, pp. 1699--1702.

\bibitem{Enkele2024}
{E. Rama, et al.}, ``{Trustworthy Integrated Circuits: From Safety to Security and Beyond},'' \emph{IEEE Access}, vol.~12, pp. 69\,603--69\,632, 2024.

\bibitem{Lopes2018}
{I. C. Lopes, et al.}, ``{Reliability analysis on case-study traffic sign convolutional neural network on APSoC},'' in \emph{2018 IEEE 19th Latin-American Test Symposium (LATS)}, 2018, pp. 1--6.

\bibitem{Du2019}
{B. Du, et al.}, ``{Ultrahigh Energy Heavy Ion Test Beam on Xilinx Kintex-7 SRAM-Based FPGA},'' \emph{IEEE Transactions on Nuclear Science}, vol.~66, no.~7, pp. 1813--1819, 2019.

\bibitem{Hoefer2023}
{J. Hoefer, et al.}, ``{SiFI-AI: A Fast and Flexible RTL Fault Simulation Framework Tailored for AI Models and Accelerators},'' in \emph{Proceedings of the Great Lakes Symposium on VLSI 2023}, ser. GLSVLSI '23.\hskip 1em plus 0.5em minus 0.4em\relax Association for Computing Machinery, 2023, pp. 287--292.

\bibitem{Sabogal2021}
{S. Sabogal, et al.}, ``{Reconfigurable Framework for Resilient Semantic Segmentation for Space Applications},'' \emph{ACM Trans. Reconfigurable Technol. Syst.}, vol.~14, no.~4, sep 2021.

\bibitem{Bohman2019}
{M. Bohman, et al.}, ``{Microcontroller Compiler-Assisted Software Fault Tolerance},'' \emph{IEEE Transactions on Nuclear Science}, vol.~66, no.~1, pp. 223--232, 2019.

\bibitem{Iturbe2016}
{X. Iturbe, et al.}, ``{A Triple Core Lock-Step (TCLS) ARM® Cortex®-R5 Processor for Safety-Critical and Ultra-Reliable Applications},'' in \emph{2016 46th Annual IEEE/IFIP International Conference on Dependable Systems and Networks Workshop (DSN-W)}, 2016, pp. 246--249.

\bibitem{de2018}
{Á. B. de Oliveira, et al.}, ``{Lockstep Dual-Core ARM A9: Implementation and Resilience Analysis Under Heavy Ion-Induced Soft Errors},'' \emph{IEEE Transactions on Nuclear Science}, vol.~65, no.~8, pp. 1783--1790, 2018.

\bibitem{Libano2019}
{F. Libano, et al.}, ``{Selective Hardening for Neural Networks in FPGAs},'' \emph{IEEE Transactions on Nuclear Science}, vol.~66, no.~1, pp. 216--222, 2019.

\bibitem{Bertoa2023}
{T. G. Bertoa, et al.}, ``{Fault-Tolerant Neural Network Accelerators With Selective TMR},'' \emph{IEEE Design \& Test}, vol.~40, no.~2, pp. 67--74, 2023.

\bibitem{Cannon2020}
{M. J. Cannon, et al.}, ``{Improving the Reliability of TMR With Nontriplicated I/O on SRAM FPGAs},'' \emph{IEEE Transactions on Nuclear Science}, vol.~67, no.~1, pp. 312--320, 2020.

\bibitem{Li2020}
{W. Li, et al.}, ``{Soft Error Mitigation for Deep Convolution Neural Network on FPGA Accelerators},'' in \emph{2020 2nd IEEE International Conference on Artificial Intelligence Circuits and Systems (AICAS)}, 2020, pp. 1--5.

\bibitem{Hari2022}
{S. K. S. Hari, et al.}, ``{Making Convolutions Resilient Via Algorithm-Based Error Detection Techniques},'' \emph{IEEE Transactions on Dependable and Secure Computing}, vol.~19, no.~4, pp. 2546--2558, 2022.

\bibitem{Younis2020}
{Y. Ibrahim, et al.}, ``{Soft errors in DNN accelerators: A comprehensive review},'' \emph{Microelectronics Reliability}, vol. 115, p. 113969, 2020.

\bibitem{Pouya2016}
P.~Taatizadeh and N.~Nicolici, ``{Automated Selection of Assertions for Bit-Flip Detection During Post-Silicon Validation},'' \emph{IEEE Transactions on Computer-Aided Design of Integrated Circuits and Systems}, vol.~35, no.~12, pp. 2118--2130, 2016.

\bibitem{Boule2007}
{M. Boule, et al.}, ``{Assertion Checkers in Verification, Silicon Debug and In-Field Diagnosis},'' in \emph{8th International Symposium on Quality Electronic Design (ISQED'07)}, 2007, pp. 613--620.

\bibitem{Tanaka2025}
{T. Tanaka, et al.}, ``{Hardware Error Detection with In-Situ Monitoring of Control Flow-Related Specifications},'' in \emph{Proceedings of Asia and South Pacific Design Automation Conference (ASP-DAC)}, 2025, pp. 966--973.

\bibitem{Zhu2020}
{Z. Zhu and B. C. Schafer}, ``{Light-Weight Soft-Errors Detection Mechanism in High-Level Synthesis},'' in \emph{2020 IEEE International Symposium on Circuits and Systems (ISCAS)}, 2020, pp. 1--5.

\bibitem{Hari2012}
{S. k. S. Hari, et al.}, ``{Low-cost program-level detectors for reducing silent data corruptions},'' in \emph{IEEE/IFIP International Conference on Dependable Systems and Networks (DSN 2012)}, 2012, pp. 1--12.

\bibitem{Didehban2024}
{M. Didehban, et al.}, ``{Generic Soft Error Data and Control Flow Error Detection by Instruction Duplication},'' \emph{IEEE Transactions on Dependable and Secure Computing}, vol.~21, no.~1, pp. 78--92, 2024.

\bibitem{Ahmadilivani2023}
{M. H. Ahmadilivani, et al.}, ``{DeepVigor: VulnerabIlity Value RanGes and FactORs for DNNs’ Reliability Assessment},'' in \emph{2023 IEEE European Test Symposium (ETS)}, 2023, pp. 1--6.

\bibitem{Chen2019}
{Z. Chen, et al.}, ``{BinFI: an efficient fault injector for safety-critical machine learning systems},'' in \emph{Proceedings of the International Conference for High Performance Computing, Networking, Storage and Analysis}, ser. SC '19.\hskip 1em plus 0.5em minus 0.4em\relax New York, NY, USA: Association for Computing Machinery, 2019.

\bibitem{Liang2014}
{L. Chen and M. Tahoori}, ``{Reliability-aware register binding for control-flow intensive designs},'' in \emph{2014 51st ACM/EDAC/IEEE Design Automation Conference (DAC)}, 2014, pp. 1--6.

\bibitem{Fleming2016}
{S. T. Fleming and D. B. Thomas}, ``{StitchUp: Automatic control flow protection for high level synthesis circuits},'' in \emph{2016 53nd ACM/EDAC/IEEE Design Automation Conference (DAC)}, 2016, pp. 1--6.

\bibitem{Nosrati2022}
{N. Nosrati, et al.}, ``{MLC: A Machine Learning Based Checker For Soft Error Detection In Embedded Processors},'' in \emph{2022 IEEE 28th International Symposium on On-Line Testing and Robust System Design (IOLTS)}, 2022, pp. 1--5.

\bibitem{Pouya2017}
P.~Taatizadeh and N.~Nicolici, ``{An automated SAT-based method for the design of on-chip bit-flip detectors},'' in \emph{2017 IEEE/ACM International Conference on Computer-Aided Design (ICCAD)}, 2017, pp. 101--108.

\bibitem{Guechi2023}
{B. Guechi, et al.}, ``{Hardware Security Module Cryptosystem Using Petri Net},'' \emph{Indonesian Journal of Electrical Engineering and Informatics (IJEEI)}, vol.~11, 06 2023.

\bibitem{Patzina2010}
{L. Patzina, et al.}, ``Monitor petri nets for security monitoring,'' in \emph{Proceedings of the International Workshop on Security and Dependability for Resource Constrained Embedded Systems}.\hskip 1em plus 0.5em minus 0.4em\relax New York, NY, USA: Association for Computing Machinery, 2010.

\bibitem{Song2021}
{S. Bai, et al.}, ``An improved petri net for fault analysis of an electronic system with hybrid fault of software and hardware,'' \emph{Engineering Failure Analysis}, vol. 120, p. 105077, 2021.

\bibitem{Wang2007}
{P. Wang, et al.}, ``{Fault Tolerance of Multiprocessor-Structured Control System by Hardware and Software Reconfiguration},'' in \emph{2007 International Conference on Mechatronics and Automation}, 2007, pp. 3745--3749.

\bibitem{Zhang2017}
{T. Zhang, et al.}, ``{Automatic Assertion Generation for Simulation, Formal Verification and Emulation},'' in \emph{2017 IEEE Computer Society Annual Symposium on VLSI (ISVLSI)}, 2017, pp. 471--476.

\bibitem{Germiniani2022}
{S. Germiniani, et al.}, ``{HARM: A Hint-Based Assertion Miner},'' \emph{IEEE Transactions on Computer-Aided Design of Integrated Circuits and Systems}, vol.~41, no.~11, pp. 4277--4288, 2022.

\bibitem{Fang2024}
{W. Fang, et al.}, ``{AssertLLM: Generating and evaluating hardware verification assertions from design specifications via multi-LLMs},'' \emph{arXiv preprint arXiv:2402.00386}, 2024.

\bibitem{Huang2022}
{M. Huang, et al.}, ``{A High Performance Multi-Bit-Width Booth Vector Systolic Accelerator for NAS Optimized Deep Learning Neural Networks},'' \emph{IEEE Transactions on Circuits and Systems I: Regular Papers}, pp. 1--13, 2022.

\bibitem{Yamawaki2018}
{A. Yamawaki, et al.}, ``{A Describing Method of An Image Processing Software in C for A High-level Synthesis Considering A Function Chaining},'' \emph{IEICE Transactions on Information and Systems}, vol. E101D, no.~2, pp. 324--334, Feb. 2018.

\bibitem{Degnan2021}
B.~Degnan, ``{Verilog Implementation of the Symmetric Block Cipher AES (NIST FIPS 197)},'' \url{https://github.com/secworks/aes}, 2021.

\bibitem{Kyle}
K.~R. Kyle Jonghyuk~Park, ``{NoC Simulator for simulating intra-chip data flow in Neural Network Accelerator},'' \url{https://github.com/KyleParkJong/Network-on-Chip-Simulator}.

\bibitem{Matsutani2009}
{H. Matsutani, et al.}, ``{Prediction router: Yet another low latency on-chip router architecture},'' in \emph{2009 IEEE 15th International Symposium on High Performance Computer Architecture}, 2009, pp. 367--378.

\bibitem{Shao2019}
{Y. S. Shao, et al.}, ``{Simba: Scaling Deep-Learning Inference with Multi-Chip-Module-Based Architecture},'' in \emph{Proceedings of the 52nd Annual IEEE/ACM International Symposium on Microarchitecture}, ser. MICRO '52.\hskip 1em plus 0.5em minus 0.4em\relax New York, NY, USA: Association for Computing Machinery, 2019, p. 14–27.

\bibitem{Shobha2010}
{S. Vasudevan, et al.}, ``Goldmine: Automatic assertion generation using data mining and static analysis,'' in \emph{2010 Design, Automation \& Test in Europe Conference \& Exhibition (DATE 2010)}, 2010, pp. 626--629.

\bibitem{Brglez1989}
{F. Brglez, et al.}, ``Combinational profiles of sequential benchmark circuits,'' in \emph{1989 IEEE International Symposium on Circuits and Systems (ISCAS)}, 1989, pp. 1929--1934 vol.3.

\end{thebibliography}

\end{document}